\documentclass[11pt]{article}
\usepackage{graphicx}
\RequirePackage{amsthm,amsmath,amsfonts,amssymb}
\RequirePackage[authoryear]{natbib}
\RequirePackage[colorlinks,citecolor=blue,urlcolor=blue]{hyperref}
\usepackage{bm}
\usepackage[pdftex,dvipsnames]{xcolor}
\usepackage[bf,font={small,sl}]{caption}
\usepackage{parskip}
\usepackage[margin=3cm]{geometry}
\usepackage{booktabs}

\usepackage{fancyhdr}
\title{\textbf{Continuous-time modelling of behavioural responses in animal movement}}
\author{Théo Michelot$^{1, 2, \dagger}$, Richard Glennie$^{1}$, Len Thomas$^{1}$,\\ Nicola Quick$^{3, 4}$, Catriona M.\ Harris$^{1}$}

\date{%
    $^1$University of St Andrews, UK\\%
    $^2$Dalhousie University, Canada\\%
    $^3$Duke University, USA\\%
    $^4$University of Plymouth, UK\\%
    $^\dagger$ \url{theo.michelot@dal.ca}
}

\begin{document}
\maketitle

\begin{abstract}
  There is great interest in ecology to understand how wild animals are affected by anthropogenic disturbances, such as sounds. For example, behavioural response studies are an important approach to quantify the impact of naval activity on marine mammals. Controlled exposure experiments are undertaken where the behaviour of animals is quantified before, during, and after exposure to a controlled sound source, often using telemetry tags (e.g., accelerometers, or satellite trackers). Statistical modelling is required to formally compare patterns before and after exposure, to quantify deviations from baseline behaviour. We propose varying-coefficient stochastic differential equations (SDEs) as a flexible framework to model such data, with two components: (1) time-varying baseline dynamics, modelled with non-parametric or random effects of time-varying covariates, and (2) a non-parametric response model, which captures deviations from baseline. SDEs are specified in continuous time, which makes it straightforward to analyse data collected at irregular time intervals, a common situation for animal tracking studies. We describe how the model can be embedded into a state-space modelling framework to account for measurement error. We present inferential methods for model fitting, model checking, and uncertainty quantification (including on the response model). We apply this approach to two behavioural response study data sets on beaked whales: a satellite track, and high-resolution depth data. Our results suggest that the whales' horizontal movement and vertical diving behaviour changed after exposure to the sound source, and future work should evaluate the severity and possible consequences of these responses. These two very different examples showcase the versatility of varying-coefficient SDEs to measure changes in behaviour, and we discuss implications of disturbances for the whales' energetic balance.
\end{abstract}

%%%%%%%%%%%%%%%%%%%%%%
%% BODY STARTS HERE %%
%%%%%%%%%%%%%%%%%%%%%%
\section{Introduction}
\label{sec:intro}

There has been a lot of effort in conservation biology to understand how human activity affects wildlife. One particular focus has been to investigate the effect of ship sonars and other anthropogenic sounds on marine mammals \citep{tyack2011, southall2019}. Controlled exposure experiments (CEEs) consist of monitoring the movement or behaviour of animals, typically using telemetry tags, before and after sound exposure, to determine if individuals respond behaviourally to the stimuli. A continuously increasing quantity of tag data now exists to address this issue, from multiple ongoing studies. This has created a need for adequate statistical methods to describe baseline behaviour and, crucially, quantify deviations from it following disturbance. A particular focus of these studies has been on identifying energetically-costly behaviours, such as sudden avoidance or interruption of foraging, as effects from these changes can accumulate to decrease animals' survival and reproductive rates. These studies have used various types of tags, with different observed variables and sampling designs, and model formulations have therefore often depended on the data type and the goal of inference.

In marine mammal studies, the most common approach to identify interruptions of foraging behaviour has been to summarise high-frequency data at the scale of dives, and compare baseline dives to exposed dives, e.g., in terms of dive duration, maximum depth, or average acceleration \citep[e.g., using DTags;][]{johnson2003}. Baseline and exposed dives can for example be compared using the Mahalanobis distance calculated using selected multivariate data streams, to identify unusual behaviour \citep{deruiter2013}. An alternative has been to use these dive summary variables to identify latent behavioural states of animals in a hidden Markov model \citep{deruiter2017}. In that context, the effect of disturbance on the probabilities of switching between the behavioural states can be estimated, to quantify the response. These approaches have usually required summarising data to the dive level, and have therefore not focused on small-scale changes in an animal's behaviour during an exposed dive. In cases where within-dive movement was analysed, the aim was to quantify whether and when a change had occurred, rather than provide a mechanistic description of the impact of disturbance on the animals' movement activity \citep[e.g.,][]{stimpert2014}.

In some studies, animals are equipped with satellite tags that record two-dimensional locations, to detect horizontal movement away from the source of disturbance \citep{cioffi2022}. Due to satellite transmission limitations, these data typically have high measurement error, and irregular intervals corresponding to times the animal came to the surface. Statistical analysis of such data is challenging, and visual assessment is typically used to measure avoidance. Recently, continuous-time discrete-space models have been proposed to analyse such noisy irregular trajectories, which require modelling animal movement on a discrete spatial grid \citep{jones2022}. % \theo{Catriona suggested mentioning that satellite tags also measure depth, and that they are also sometimes used to look into vertical response. I don't think that it's necessary to discuss this here, as it's nicer to clearly distinguish between DTags and Argos for our purposes.}

We propose varying-coefficient stochastic differential equations (SDEs) as a versatile method to estimate behavioural responses from different types of CEE tag data. Multiple SDE formulations have been proposed for the analysis of animal movement data, including Brownian motion \citep{pozdnyakov2014}, Ornstein-Uhlenbeck processes \citep{dunn1977}, and the integrated Ornstein-Uhlenbeck process \citep{johnson2008}. In those models, the animal's movement dynamics are specified in terms of a few parameters, e.g., representing mean speed or autocorrelation. The varying-coefficient approach we propose here provides great flexibility to express these parameters as functions of time-varying covariates \citep{michelot2021}. We demonstrate the utility of this approach for behavioural response studies, based on several extensions to the approach of \cite{michelot2021}: (1) estimation of deviations from baseline model using difference smooths, (2) uncertainty quantification using simultaneous intervals, (3) measurement error using state-space models, and (4) model checking using posterior predictive checks. We illustrate the utility of these models with two common types of CEE data: high-resolution data on diving behaviour from archival tags, and low-resolution position data from satellite tags. This statistical framework is widely applicable beyond these examples.

\section{Beaked whale movement data}
\label{sec:data}

Beaked whales have been the focus of multiple behavioural response studies (BRS) due to their apparent vulnerability to the effects of military sonar systems (DeRuiter et al 2013, Southall et al. 2016; Tyack et al. 2011). For this purpose, CEEs have been conducted with different types of animal-borne tags, including movement and acoustic sensors at different resolutions, to detect individual behavioral response (Southall et al., 2016). In this paper, we focus on CEEs for Cuvier's beaked whales (\emph{Ziphius cavirostris}), and analyse two types of data with different variables and resolutions: coarse two-dimensional location data from a satellite tag, and fine-resolution depth data from DTags. Plots of the data are shown in Appendix A.

\subsection{Satellite tag data}
\label{sec:data_sat}

The satellite tag analysed here was deployed as part of the Atlantic BRS, a study on the effects of mid-frequency active sonar on deep diving whales. The tag was a SPLASH10-292, Argos satellite-linked location-depth tag (produced by Wildlife Computers, Redmond, Washington) remotely deployed using a DAN-INJECT JM 25 pneumatic projector (DanWild LLC, Austin, Texas) in the LIMPET configuration \citep{andrews2008} from a 9m rigid-hulled aluminium boat.

The tag was deployed on an adult male Cuvier’s beaked whale off Cape Hatteras, North Carolina, on the 24th May 2018, and it transmitted for 38 days. Location estimates were derived from Service Argos receivers on polar-orbiting satellites, and were assigned an accuracy class based on the timing and number of transmissions received during a satellite pass \citep[see][for details]{foley2021}. Only the higher accuracy positions were used. Reliable locations can only be recorded when the whale is at the surface, and when satellites are available, which severely limits data collection. Specifically, the latitude of the study site provides only 9\% temporal satellite coverage \citep{cioffi2022}, and the whales spend most of their time deep underwater, with average 2.2 minute surface ventilation periods \citep{shearer2019}. As a result, the locations were sparse in time (average of 2 locations per day), and included measurement error. The measurement error was available in the form of error ellipses, each corresponding to the $\sqrt{2}$-sigma contour of a bivariate normal distribution \citep{mcclintock2015}.

To increase data resolution, an Argos goniometer (Woods Hole Group Inc., Bourne, MA, USA) was deployed from the research vessel to collect further data from the tagged whale’s transmitter \citep{cioffi2022}. Locations from the goniometer had high spatiotemporal resolution, but they only covered short time periods when the vessel was within range of the whale. We added these data to the satellite trajectory to increase the information available for this analysis. We assumed that the goniometer locations had isotropic error ellipses (as defined above), with radius that depended on the strength of the signal received (a proxy for distance between the vessel and the whale). Specifically, we set the radius to 100 m when the signal was stronger than -50 dB, 500 m between -51 and -70 dB, 1 km between -71 and -80 dB, 2 km between -81 and 90 dB, and 10 km for signals weaker than -91 dB.

On June 3rd 2018, at 16:00:04 UTC, the whale was exposed to an hour-long CEE of mid-frequency active sonar, similar to the tactical sonars used by the US and other navies \citep{southall2016experimental}. Data visualisation suggests that the whale moved away from the sonar source \citep{southall2020}, but this has not been confirmed by statistical analysis.

% \begin{itemize}
% \item measurement error
% \item irregular time intervals
% \item long time period (a few weeks)
% \item description of the zc69 record, which is what we analysed
% \item mention that these are supplemented by goniometer data, which are the opposite (high-res, short-term)
% \end{itemize}

\subsection{DTag data}
\label{sec:data_dtag}

DTAGs are multi-sensor archival tags that are attached to animals via suction cups for up to tens of hours, and record various acoustic and movement variables, including depth at 50Hz resolution \citep{johnson2003}. Our analysis included data from two separate studies: the SOCAL BRS \citep[four tags; for full tag details, see][]{deruiter2013, southall2016experimental}, and the Atlantic BRS \citep[one tag; for details, see][]{southall2020}. The tags were programmed to release after a predetermined period, if they had not already detached from the animal, and were recovered to download recorded data. Pressure recordings were converted to depths and orientation offset from tag position were performed using calibration information for each tag \citep{johnson2003}. Data were downsampled to 15-sec resolution for analysis to reduce computational effort.

Beaked whales typically perform two types of dives: deep dives (up to several kilometers of depth), during which their foraging activity occurs, and shallow dives \citep{shearer2019}. For this analysis, we only retained dives with a maximum depth greater than 700m, to investigate changes in foraging behaviour \citep{deruiter2013}. The processed data set included 13 dives, each about 1 hour in length.

Two of the whales were exposed to mid-frequency active sonar during a deep dive, each for a period of 30 min. \cite{deruiter2013} used dive summaries, such as duration and maximum depth, to investigate behavioural changes following sonar exposure from these data. In this paper, we propose a different approach based directly on the high-resolution data, which focuses on response at a short temporal scale after the start of exposure.

% \begin{itemize}
% \item ``behaviour'' tag rather than movement tag per se
% \item high temporal resolution
% \item various variables: accelerometry, body posture, depth
% \item short time period (a few hours)
% \item description of the records analysed in this paper (zc10\_272, zc11\_267, zc13\_210, zc13\_211, zc17\_234)
% \item maybe explain that we focus on deep dives
% \end{itemize}

\section{Behavioural response model}
\label{sec:model}

One approach to describing behavioural responses of animals consists in specifying two components: a model of baseline behaviour, and a model for deviations from that baseline (``response'' behaviour). The mathematical formulation of both components will depend on the specific application, and should be informed by the research question. One key point is that the definition of the baseline model partially (and implicitly) determines what constitutes a behavioural response, regardless of the model used for the deviations. In this section, we propose varying-coefficient SDEs as a flexible model of baseline behaviour, applicable to various data types. We describe how responses can be modelled in that framework, in particular using difference smooths, and we discuss underlying assumptions. 

\subsection{Varying-coefficient stochastic differential equations}
\label{sec:model1}

Varying-coefficient SDEs are a versatile class of time series models with time-varying dynamics \citep{michelot2021}. We consider the Itô SDE for the continuous-time process $(Z_t)$,
\begin{equation*}
  dZ_t = \mu(Z_t, \bm\theta_t) \mathop{dt} + \sigma (Z_t, \bm\theta_t) \mathop{dW_t},
\end{equation*}
where $\mu$ is the drift function and $\sigma$ the diffusion function, $W_t$ is a standard Wiener process, and $\bm\theta_t$ is a vector of time-varying parameters. The drift $\mu$ measures the expected change over infinitesimal time increments, e.g., $\mu$ might capture the preferred direction of movement if $Z_t$ is the location or velocity of an animal. The diffusion $\sigma$ captures stochastic variability around this expected change.

The functions $\mu$ and $\sigma$ are often chosen to have a simple parametric form, to help with implementation and interpretation. In the following, we use Brownian motion and the Ornstein-Uhlenbeck process for illustration, as these are the models we use in the case study, but the methodology generalises to other SDEs. In the case of Brownian motion, we have $\mu(Z_t, \bm\theta_t) = a$ and $\sigma(Z_t, \bm\theta_t) = \sigma$, where $a \in \mathbb{R}$ and $\sigma > 0$ are constant drift and diffusion parameters, respectively. Similarly, the Ornstein-Uhlenbeck process is defined by the SDE with $\mu(Z_t, \bm\theta_t) = b (a - Z_t)$ and $\sigma(Z_t, \bm\theta_t) = \sigma$, where $a \in \mathbb{R}$ is the long-term mean of the process, $b > 0$ is the strength of the attraction to the mean, and $\sigma > 0$ measures the volatility. In the varying-coefficient approach, the parameters of the SDE are specified as time-varying functions of covariates. This allows for great flexibility in the dynamics of the modelled process, while retaining the simple interpretation of parametric SDEs.

The derivation of the likelihood of an SDE observed at discrete time intervals requires evaluating its transition density, i.e., the function $p(Z_{t+\Delta} \mid Z_t)$ for each time interval of observation $\Delta > 0$. This transition density is known in closed form for the special cases considered here (Brownian motion and Ornstein-Uhlenbeck process) and, in the varying-coefficient setting, we use the value of the parameter at the start of the interval. This is an approximation based on the assumption that the SDE parameters are constant over the time interval of observation. That is, we use the transition densities
\begin{align*}
  \text{Brownian motion:}\quad & Z_{t_1} \mid Z_{t_0} = z_0 \sim N \left[ z_0 + a_{t_0} \Delta,\ \sigma_{t_0}^2 \Delta \right] \\
  \text{Ornstein-Uhlenbeck:}\quad & Z_{t_1} \mid Z_{t_0} = z_0 \sim N \left[ (1 - e^{-b_{t_0}\Delta}) a_{t_0} + e^{-b_{t_0}\Delta} z_0,\ \frac{\sigma_{t_0}^2}{2b_{t_0}} (1 - e^{-2b_{t_0}\Delta}) \right]
\end{align*}
where $\Delta = t_1 - t_0$. More generally, when the transition density is not tractable, a discretisation approach such as the Euler-Maruyama method can be used \citep{michelot2021}.

We use the formalism of generalised additive models (GAMs) to specify each parameter $\theta_t$ as a function of covariates,
\begin{equation}
  \label{eqn:gam}
  h(\theta_t) = \alpha_0 + f_1(x_{1t}) + f_2(x_{2t}) + \dots,
\end{equation}
where $h$ is a link function, $\alpha_0$ is an intercept parameter, and the function $f_j$ represents the relationship between the covariate $x_{j}$ and the parameter. We model the functions $f_j$ using penalised splines, which can capture linear or non-linear relationships, as well as random effects \citep{michelot2021}. The goal of inference is then to estimate linear model components $\bm\alpha$ (e.g., the intercept), basis function coefficients $\bm\beta$ for the non-parametric relationships, and smoothness parameters $\bm\lambda$ of non-linear functions (or precision of random effects). We view this as a mixed effect model, where the basis coefficients $\bm\beta$ are treated as random effects. The marginal likelihood of such a model, where the random effects have been integrated out, can be computed using the Laplace approximation, and we implemented it using the Template Model Builder (TMB) R package \citep{kristensen2015}.

In a few simple special cases, this model reduces to a GAM, or to a GAM for location, scale and shape \citep[GAMLSS;][]{rigby2005}. In particular, the Brownian motion described above can be written as $(Z_{t_1} - Z_{t_0})/\Delta \sim N \left[ a_{t_0},\ \sigma_{t_0}^2 \right]$. This is a GAMLSS with response variable $(Z_{t_1} - Z_{t_0})/\Delta$, where the response distribution is normal, and the mean (i.e., location) and standard deviation (i.e., scale) are modelled as non-parametric functions of covariates. As a consequence, GAMLSS software such as the gamlss package in R can be used directly in this case \citep{stasinopoulos2008}, but this does not apply to general SDEs (e.g., the Ornstein-Uhlenbeck process).

SDEs have been used to model various types of animal behaviours, including movement around a central location \citep{dunn1977}, highly directional movement \citep{johnson2008}, and habitat selection \citep{michelot2019langevin}. Their continuous-time formulation makes it possible to analyse irregularly-sampled data, and to compare or combine studies with different sampling schemes. In this framework, the specification of a baseline model requires the choice of: (1) an appropriate SDE, informed by the type of data and the animal's movement patterns, and (2) relevant covariates to be included in the SDE parameters. The SDE should capture a template of the animal's behaviour under normal conditions, so that deviations from that template can be quantified. We describe two examples in Section \ref{sec:model3}.

\subsection{Modelling the response}
\label{sec:model2}

In many animal movement analyses, it is of interest to detect behavioural changes, or to compare behaviour over different phases of data. This is particularly relevant to identify effects of internal or external influences on behaviour, such as anthropogenic disturbance or habitat degradation.

In the framework of varying-coefficient SDEs, we propose decomposing the animal's movement parameters into different components for baseline and response behaviours. This can be modelled within the additive structure of Equation \ref{eqn:gam}, where time-varying terms can be included to capture behavioural changes after disturbance. The form of these additional terms, and the choice of the parameter on which to include them,  will generally depend on the application, as different formulations might be required for different types of deviations from baseline. Perhaps the simplest response model would be to add an intercept term during the sound exposure (or for some set period after start of exposure). This could for example capture an unusually high (or low) level of activity directly following disturbance. This simple model requires specifying a time period over which to include the additional intercept and, although this choice could be based on biological expertise, it might be difficult in many applications. Alternatively, in studies where the level of disturbance is measured (e.g., received sound level), this could directly be included as a covariate acting on the SDE parameters. This option is attractive due to its mechanistic interpretation, but direct measurements of disturbance are not always available.

\begin{figure}[htbp]
  \centering
  \includegraphics[width=\textwidth]{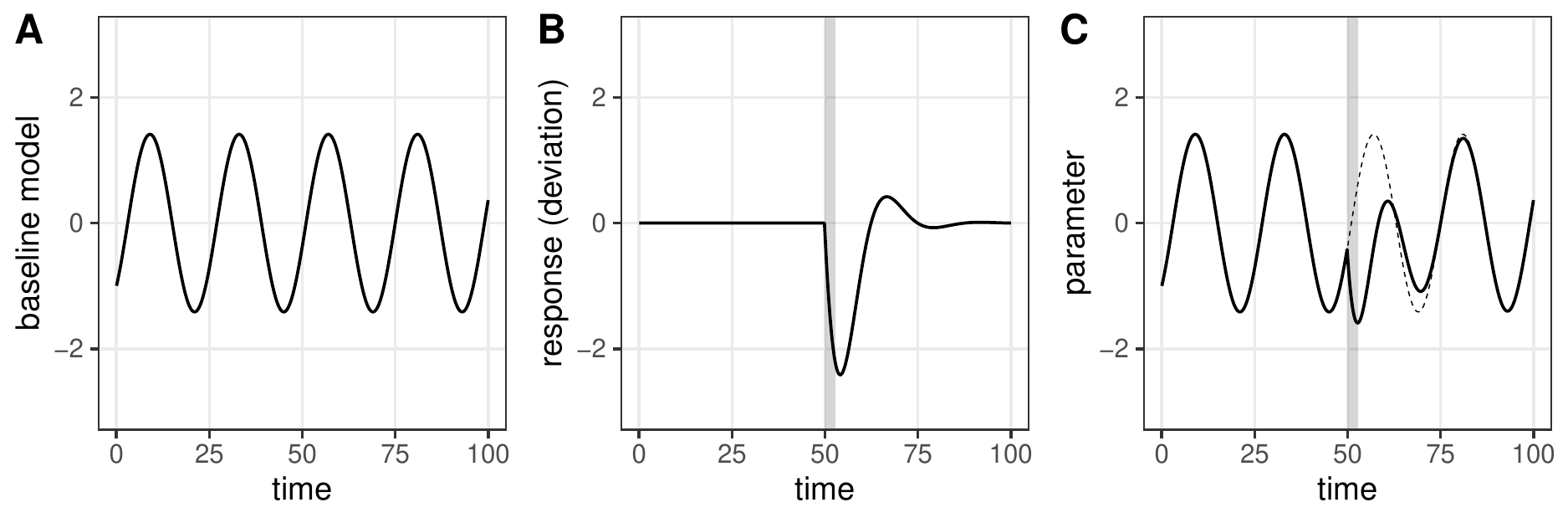}
  \caption{Example model terms plotted over time: (A) baseline covariate effect $f_1(x_{1t})$, e.g., of time of day; (B) difference smooth measuring deviation from baseline, $f_d(x_{d,t})$, which eventually decays to zero; and (B) SDE parameter $\theta_t$, obtained as $\theta_t = f_1(x_{1t}) + f_d(x_{d, t})$. The shaded band shows the period of exposure, and the dashed line in (C) is the value of the parameter in the baseline model. In a behavioural response study, the main focus is the deviation from baseline (B).}
  \label{fig:spline_illu}
\end{figure}

In the following, we propose using separate smooth relationships between parameters and covariates for the baseline and response phases of the data. More specifically, we suggest estimating one smooth function for baseline, and a ``difference smooth'' to measure the discrepancy between baseline and post-disturbance periods. Figure \ref{fig:spline_illu} illustrates the concept of a difference smooth, and we provide details for the model formulations of interest in the next section. The main assumption of this approach is that the deviation between baseline and response behaviours can be modelled using a smooth function. This would for example be violated if the response of interest is a ``jerk'' reaction corresponding to a large, yet momentary, change in movement dynamics. In cases where the assumption of smoothness holds, however, difference smooths are a convenient formulation, as they make it possible to directly carry out inference (including uncertainty quantification) on the discrepancy between baseline and post-disturbance behaviour.

\subsection{Case studies}
\label{sec:model3}

\subsubsection{Horizontal avoidance}
\label{sec:methods_horizontal}

Avoidance is defined as movement away from a disturbance, and it has been documented in beaked whales \citep[e.g.,][]{tyack2011}. When this behaviour is observed in the Easting-Northing plane (rather than in the depth dimension), we call it horizontal avoidance. In the framework of varying-coefficient SDEs, we propose modelling horizontal avoidance as follows. We define the location process $(\bm{Z}_t)$ of the animal (Easting-Northing) as an isotropic two-dimensional Ornstein-Uhlenbeck process, where each coordinate is defined by
\begin{equation}
  \label{eqn:sde_ou}
  dZ_t = b (a_t - Z_t) \mathop{dt} + \sigma \mathop{dW_t},
\end{equation}
where $a_t$ is a time-varying centre of attraction, $b > 0$ is the strength of attraction to $a$, and $\sigma > 0$ is the diffusion parameter. We define the time-varying centre of attraction as
\begin{equation}
  \label{eqn:mu_ou}
  a_t = \alpha_0 + f_d(t) I_{\{t \geq t_\text{exp}\}},
\end{equation}
where $I$ is the indicator function, and $t_\text{exp}$ is the time of start of exposure. Combining Equations \ref{eqn:sde_ou} and \ref{eqn:mu_ou}, we obtain the following SDEs before and after start of exposure,
\begin{align*}
  \text{Before:}\quad & dZ_t = b (\alpha_0 - Z_t) \mathop{dt} + \sigma \mathop{dW_t}\\
  \text{After:}\quad & dZ_t = b (\alpha_0 - Z_t) \mathop{dt} + b f_d(t) dt + \sigma \mathop{dW_t}.
\end{align*}
The ``before'' model is an OU process with centre of attraction $\alpha_0$. The ``after'' model is a modification of that process with an additional drift (or ``advection'') term $b f_d(t)$. That is, the animal's movement after start of exposure is driven by two opposing forces: attraction towards a long-term central location $\alpha_0$, and time-varying advection away from that location, corresponding to deviation from baseline. In this example, the function $f_d$ can therefore be interpreted as the animal's horizontal avoidance.

\subsubsection{Disruption of foraging dive behaviour}
\label{sec:methods_vertical}

Past studies have reported that sound exposure can prompt a beaked whale to stop echolocating during a foraging deep dive, leading to decreased potential energetic gains \citep[e.g.,][]{deruiter2013}. Beaked whales have also been observed to extend non-foraging dives to depths beyond that of non-foraging dives observed in baseline, which may increase energetic costs. Here, we model the depth $D_t$ using varying-coefficient Brownian motion,
\begin{equation}
  \label{eqn:sde_bm}
  dD_t = a_t \mathop{dt} + \sigma_t \mathop{dW_t},
\end{equation}
where $a_t$ and $\sigma_t$ are the time-varying drift and diffusion parameters, respectively. The drift is modelled as a function of proportion of time through dive $x_{1t} \in [0, 1]$, to capture the shape of dives,
\begin{equation}
  \label{eqn:par_bm1}
  a_t = \alpha_0^a + f_1^a(x_{1t}).
\end{equation}

The diffusion $\sigma_t$ is also assumed to depend on $x_{1t}$, with a different relationship during baseline and response phases. We also include a dive-specific random intercept in $\sigma_t$ to capture heterogeneity in the data. Finally, the model is
\begin{equation}
  \label{eqn:par_bm2}
  \log( \sigma_t ) = \alpha_0^\sigma + \alpha_{d_t}^\sigma + f_B^\sigma(x_{1t}) + \sum_{k = 1}^K f_{R,k}^\sigma(x_{1t}) \times I_{\{d_t = k\}} \times I_{\{x_{2t} = 1\}}
\end{equation}
where $d_t \in \{ 1, \dots, K\}$ is the dive index at time $t$, $f_B^\sigma$ describes the baseline model, $f_{R, k}^\sigma$ is the difference smooth for exposed dive $k$, and $x_{2t}$ is a binary variable equal to 0 before start of exposure and 1 after. The parameter $\alpha_0^\sigma$ is the population-level mean intercept, and the dive-specific random intercepts are assumed to follow $\alpha_k^\sigma \stackrel{i.i.d.}{\sim} N(0, \nu^2)$, where $\nu^2$ measures variance around the population mean. The indicator functions ensure that a separate difference smooth is included for each exposed dive, and that it is only added after the start of exposure.

\subsection{Measurement error using state-space models}
\label{sec:SSM}

\subsubsection{State-space formulation}

Measurement error is common in animal tracking data, and it is in particular present in the Argos locations analysed in Section \ref{sec:whales}. State-space models have been proposed to account for observation error in animal movement studies \citep{anderson1991, jonsen2003}. In this section, we describe how the SDEs presented above can be embedded into a state-space formulation. In the case study, measurement error only arises in two-dimensional tracking data used to detect horizontal avoidance. Therefore, we present the methods in the special case of the varying-coefficient Ornstein-Uhlenbeck process (described in Section \ref{sec:methods_horizontal}). However, the approach can be applied directly to other SDEs where the transition density is normal (or approximately normal, e.g., under the Euler-Maruyama discretisation).

Let $\bm{Z}_t$ be the two-dimensional position of the animal at time $t$, described by an isotropic Ornstein-Uhlenbeck process (i.e., both dimensions are described by the same parameters), and let $\tilde{\bm{Z}}_i$ be a (noisy) observation obtained at time $t_i$. Assuming that the measurement error can be modelled with a normal distribution, we consider the state-space formulation with the following observation and latent state equations,
\begin{align}
  & \text{Observation:}\quad \tilde{\bm{Z}}_i = \bm{Z}_{t_i} + \bm\varepsilon_i,\quad \bm\varepsilon_i \sim N(\bm{0}, \bm\Omega_i) \label{eqn:ssm1}\\
  & \text{Latent state:}\quad \bm{Z}_{t_{i+1}} \sim N \left[ \left( 1 - e^{-b_{t_i}\Delta_i} \right) \bm{a}_{t_i} + e^{-b_{t_i}\Delta_i} \bm{Z}_{t_i},\ \frac{\sigma_{t_i}^2}{2b_{t_i}} \left( 1 - e^{-2b_{t_i}\Delta_i} \right) \bm{I}_2 \right] \label{eqn:ssm2}
\end{align}
where $\Delta_i = t_{i+1} - t_i$, $\bm\Omega_i$ is the measurement error covariance matrix at time $t_i$, and $\bm{I}_2$ is the 2$\times$2 identity matrix. Here, the latent state equation is simply the transition density of the Ornstein-Uhlenbeck process.

\subsubsection{Inference with the Kalman filter}

Inference for this model can be carried out using the Kalman filter, which provides a computationally efficient method to evaluate the likelihood and to obtain one-step-ahead estimates of the latent state variables \citep{durbin2012}. To apply this method, we rewrite Equations \ref{eqn:ssm1} and \ref{eqn:ssm2} as linear equations in matrix notation. The observation model (Equation \ref{eqn:ssm1}) can be written as
\begin{equation}
  \label{eq:SSM_obs}
  \tilde{\bm{Z}}_i = \bm{A} \bm{Z}_{t_i} + \bm\varepsilon_i,\quad \bm\varepsilon_i \sim N(\bm{0}, \bm\Omega_i) 
\end{equation}
where $\bm{A} = \bm{I}_2$. Similarly, we can write the latent state model (Equation \ref{eqn:ssm2}) as
\begin{equation*}
  \bm{Z}_{t_{i+1}} = \bm{T}_i \bm{Z}_{t_i} + \bm{B}_i \bm{u}_i + \bm\eta_i,\quad \bm\eta_i \sim N(\bm{0}, \bm{Q_i})
\end{equation*}
where $\bm{T}_i  = e^{-b_{t_i}\Delta_i} \bm{I}_2$, $\bm{B}_i = (1 - e^{-b_{t_i}\Delta_i}) \bm{I}_2$, $\bm{u}_i = \bm{a}_{t_i}$, and $\bm{Q}_i = \sigma_{t_i}^2 (1 - e^{-2b_{t_i}\Delta_i}) / (2b_{t_i}) \bm{I}_2$. \cite{durbin2012} describe the algorithm for the Kalman filter in terms of those matrices (Section 4.3.2), and the derivation of the model likelihood as a by-product (Section 7.2.1).

Other state-space model methods can be applied directly using the formulation highlighted above, such as Kalman smoothing. Kalman smoothing is an algorithm that returns predictions of the latent state variables given the full observed time series, as well as uncertainty estimates \citep{durbin2012}. This might be particularly useful in studies where reconstructing the true trajectory of an animal is of primary interest.

\section{Uncertainty quantification and model checking}

\subsection{Confidence intervals for non-parametric terms}

A key challenge is to estimate the difference smooths and determine whether they clearly deviate from zero. One approach is to compute confidence intervals on the difference smooths, where overlap with zero may be interpreted as lack of clear deviation. Two types of confidence intervals can be derived for a smooth function, with different interpretations. \emph{Pointwise} confidence intervals can only be used to make statements about uncertainty at a given covariate value, whereas \emph{simultaneous} confidence intervals represent joint uncertainty across the domain of definition of the function. Beyond difference smooths, confidence intervals are crucial to interpret relationships between SDE parameters and covariates (e.g., in baseline model).

The two types of intervals are contrasted in Figure \ref{fig:CI}, and we describe methods to derive them based on posterior simulations. Note that, even though we do not carry out full Bayesian inference, we use terminology from the empirical Bayes view of hierarchical models \citep{miller2019}. We therefore call the joint distribution of fixed and random effect parameters the ``posterior''. This is approximated by a multivariate normal distribution centred on the maximum likelihood estimates $\hat{\bm\gamma} = (\hat{\bm\alpha}, \hat{\bm\beta})$, with covariance matrix $\hat{\bm\Sigma}$ derived from the inverse of the Hessian of the log-likelihood \citep[e.g., using the function \texttt{sdreport} in the TMB package;][]{kristensen2015}.

\subsubsection{Pointwise confidence intervals}

Consider a grid over the range of the covariate of interest, $(x_1, x_2, \dots, x_M)$. We can obtain pointwise $100(1 - \alpha)\%$ confidence intervals, as follows:
\begin{enumerate}
\item Generate $K$ posterior draws of all fixed and random effect parameters from $N(\hat{\bm\gamma}, \hat{\bm\Sigma})$.
\item From these posterior draws, derive $K$ realisations of the smooth function.
\item For each point $x_m$ of the grid, compute quantiles of the $K$ functions with probabilities $\alpha/2$ and $1 - \alpha / 2$. These correspond to the lower and upper bounds of the confidence interval, respectively.
\end{enumerate}

\begin{figure}[htbp]
  \centering
  \includegraphics[width=\textwidth]{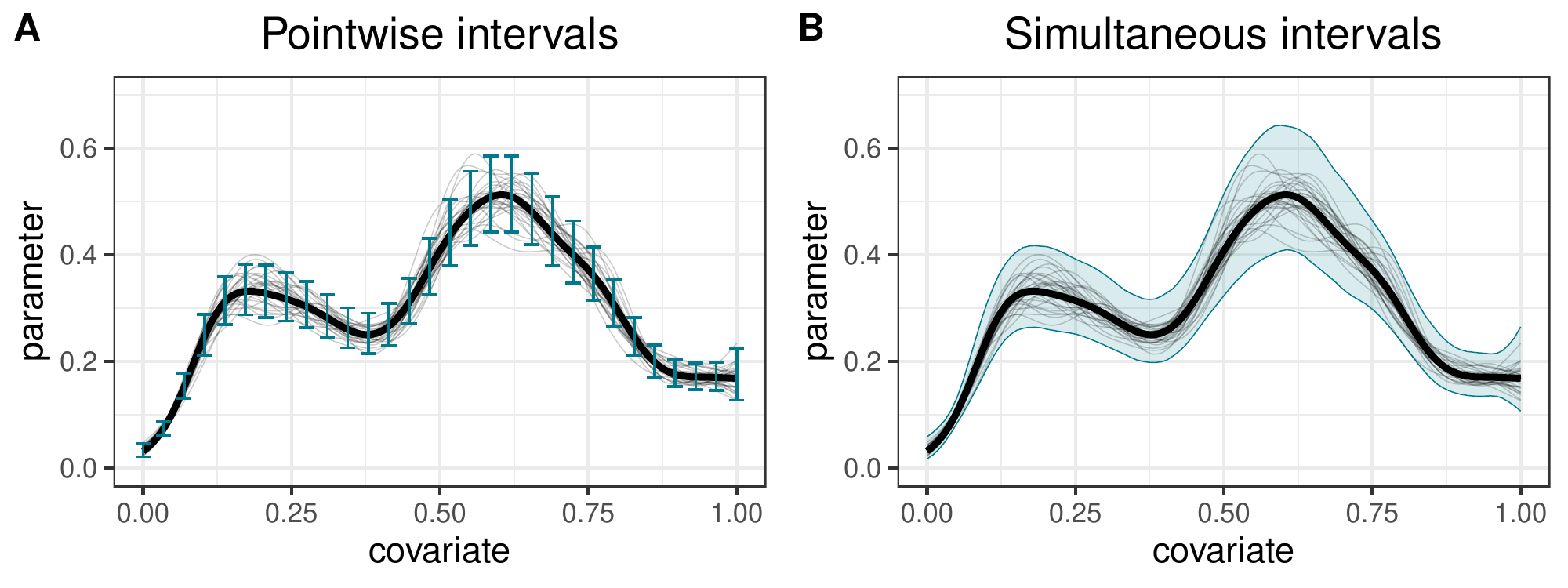}
  \caption{Illustration of confidence intervals. The thick black line is the mean estimate, and thin black lines are posterior samples for the spline. (A) The vertical segments show pointwise 95\% confidence intervals on a grid of values of the covariate. (B) The shaded area is a 95\% simultaneous confidence band.}
  \label{fig:CI}
\end{figure}

From a Bayesian viewpoint, the $K$ realisations are draws from the posterior distribution of the smooth, and the interval can therefore be interpreted as a credible interval. They can also be viewed as confidence intervals, with the expected coverage ``across the function'' \citep{marra2012}. That is, if we denote as $p_m$ the proportion of such intervals that include the true function at $x_m$, then we expect $(p_1 + p_2 + \dots p_M )/M \approx 1- \alpha$ (e.g., 0.95 for 95\% confidence intervals). Pointwise confidence intervals are illustrated in Figure \ref{fig:CI}(A).

Although this procedure is for one smooth function, confidence intervals on the SDE parameter $\theta_t$ can be derived similarly. The only modification is that, in step 2, a realisation of the SDE parameter across the covariate grid needs to be computed. This requires adding other model terms (i.e., effects of other covariates, fixed to a given value), and applying the inverse link function.

\subsubsection{Simultaneous confidence intervals}

An alternative approach to quantify uncertainty in a function is to derive simultaneous confidence intervals. A simultaneous $100(1 - \alpha)\%$ confidence band has the following interpretation: $100(1 - \alpha)\%$ of such confidence intervals will include the true smooth function in its entirety. This requirement is more stringent than for pointwise intervals, and simultaneous intervals therefore tend to be wider. Figure \ref{fig:CI}(B) shows an example of 95\% simultaneous confidence band.

Here, we follow the simulation-based method described by \cite{ruppert2003} to obtain simultaneous confidence intervals over the grid $\bm{x} = (x_1, \dots, x_M)$. We outline the main steps, but refer to Section 6.5 of \cite{ruppert2003} for details. In this section, we denote as $f$ the true function, $\hat{f}$ the estimated smooth, $\bm{y} = (f(x_1), \dots, f(x_M))$, and $\hat{\bm{y}} = (\hat{f}(x_1), \dots, \hat{f}(x_M))$. 

\begin{enumerate}
\item Generate $K$ posterior draws of $\hat{\bm\gamma} - \bm\gamma$ from $N(\bm{0}, \hat{\bm\Sigma})$.
\item From each posterior draw, derive a realisation of the difference between the true function and the estimated smooth, as $\hat{\bm{y}} - \bm{y} = \bm{C}_x (\hat{\bm\gamma} - \bm\gamma)$, where $\bm{C}_x$ is the design matrix of basis function evaluations over $\bm{x}$.
\item From each realisation, approximate the standardised difference between $\hat{f}$ and $f$ by
  \begin{equation*}
    H = \max_{m = 1, \dots, M} \left| \frac{(\bm{C}_x [\hat{\bm\gamma} - \bm\gamma] )_m}{\widehat{\text{SD}}(\hat{y}_m - y_m)} \right|,
  \end{equation*}
  where the standard deviation in the denominator is measured from the $K$ replications of $\hat{\bm{y}} - \bm{y}$.
\item The simultaneous confidence interval is
  \begin{equation*}
    \hat{\bm{y}} \pm q_{1 - \alpha} \widehat{\text{SD}}(\hat{\bm{y}} - \bm{y}),
  \end{equation*}
  where $q_{1-\alpha}$ is the $(1 - \alpha)$ quantile of $H$.
\end{enumerate}

% \theo{I wonder if the notation in the above equations could be further simplified, e.g., replacing numerator in $H$ by $\hat{y}_m - y_m$. Why didn't Ruppert et al do this?}

The choice between pointwise and simultaneous intervals depends on the application, and on whether joint statements across the range of the smooth are required. We suggest that simultaneous confidence intervals are a natural choice to quantify uncertainty on the difference smooths that measure deviations from baseline behaviour (e.g., $f_d$ in Equation \ref{eqn:mu_ou}). Indeed, to determine whether there is clear evidence of deviation, the question of interest is whether the identically zero function is included in the confidence region (rather than whether the confidence region overlaps zero for some covariate value, which is a weaker statement). As with pointwise intervals, confidence bands for the SDE parameter can also be computed using this method, where additional model terms need to be added to $\bm{C}_x$.

\subsection{Posterior predictive checks}
\label{sec:check}

We propose a simulation-based approach to model checking for a fitted varying-coefficient SDE. The general idea is to simulate from the fitted model, and compare patterns in the simulated data and in the observed data, where discrepancies suggest lack of fit. The suggested procedure is as follows:
\begin{enumerate}
\item Generate $K$ draws from the posterior distribution of fixed and random effects, $\{ \bm\gamma^{(1)}, \dots, \bm\gamma^{(K)} \}$. 
\item Using each posterior draw $\bm\gamma^{(k)}$, simulate a time series $\bm{z}^{(k)}$ over an appropriate time period for comparison with the observed time series.
\item Compute a relevant summary statistic for each simulated time series, $g(\bm{z}^{(1)}), \dots, g(\bm{z}^{(K)})$, which measures an important feature of the data-generating process.
\item Compute the summary statistic for the observed time series, $g(\bm{z})$.
\item Compare $g(\bm{z})$ to the distribution of the $g(\bm{z}^{(k)})$, to assess how compatible the observed data are with the estimated model. This could for example involve the computation of a $p$-value, as the proportion of $g(\bm{z}^{(k)})$ which are more extreme than $g(\bm{z})$.
\end{enumerate}

This method is used to check goodness-of-fit of a model of baseline diving behaviour in Section \ref{sec:vertical}, where the test statistics are characteristics of a dive (e.g., proportion of time descending, proportion of time spent under 500m).

% \subsection{Further extensions}

% Hopefully, the point should be that these extensions are very powerful mathematically, but don't require a lot of effort to implement because the framework is so flexible.

% \subsubsection{Hierarchical modelling}

% (e.g., useful in cases where there are no clear dive types, or for multiple species)

% The flexibility of the spline-based formulation proposed in Section \ref{sec:model1} can be further extended using ideas from hierarchical modelling \citep{pedersen2019}, to allow for additional variability between groups of data (e.g., between animals, between tracks, between species). The simplest option, to allow for heterogeneity between groups, is to include iid normal random intercepts. This is straightforward in this framework, as such a random effect can be viewed as a special case of basis-penalty smooth [REFERENCE].

% In cases where a random intercept is not sufficient to capture the heterogeneity, the shape and smoothness of the splines can be specified as group-dependent. 

% \subsubsection{Smooth term derivatives}

% Could the derivative of the smooth terms be useful? They are tractable, and maybe could be used to find change points. Gavin Simpson's blog has stuff about this.

\section{Simulation study}
\label{sec:sim}

We assessed the performance of the workflow outlined in Section \ref{sec:model1} to estimate deviations from baseline using simulations. We simulated data from a Brownian motion, with drift and diffusion parameters specified as known functions of a time-varying covariate $x_1$. The relationship between the diffusion parameter and $x_1$ took two different forms, depending on a binary covariate representing disturbance or response behaviour. We then fitted a varying-coefficient SDE, with a difference smooth to capture the discrepancy in diffusion between baseline and response behaviour.

We simulated data from a Brownian motion with no drift and with time-varying diffusion parameter defined as,
\begin{align*}
  \text{Diffusion (baseline):}\quad & \sigma_t^{B} = 0.5 - 1.5 (x_{1t} - 0.5)^2\\
  \text{Diffusion (response):}\quad & \sigma_t^{R} = 0.05 + 5 (x_{1t} - 0.5)^2
\end{align*}
where $x_{1t} \in [0, 1]$ was analogous to ``proportion of time through dive'' in the beaked whale diving study (Section \ref{sec:methods_vertical}). For each iteration of the simulation, we generated nine independent time series: eight from the baseline model, and one that started in the baseline model and switched to the response model when $x_{1t} \geq 0.25$. Each time series contained $n = 200$ points, at random times uniformly distributed between $t_1 = 0$ and $t_n = 10$, to check that the method works with irregular time intervals.

\begin{figure}[htbp]
  \centering
  \includegraphics[width=0.49\textwidth]{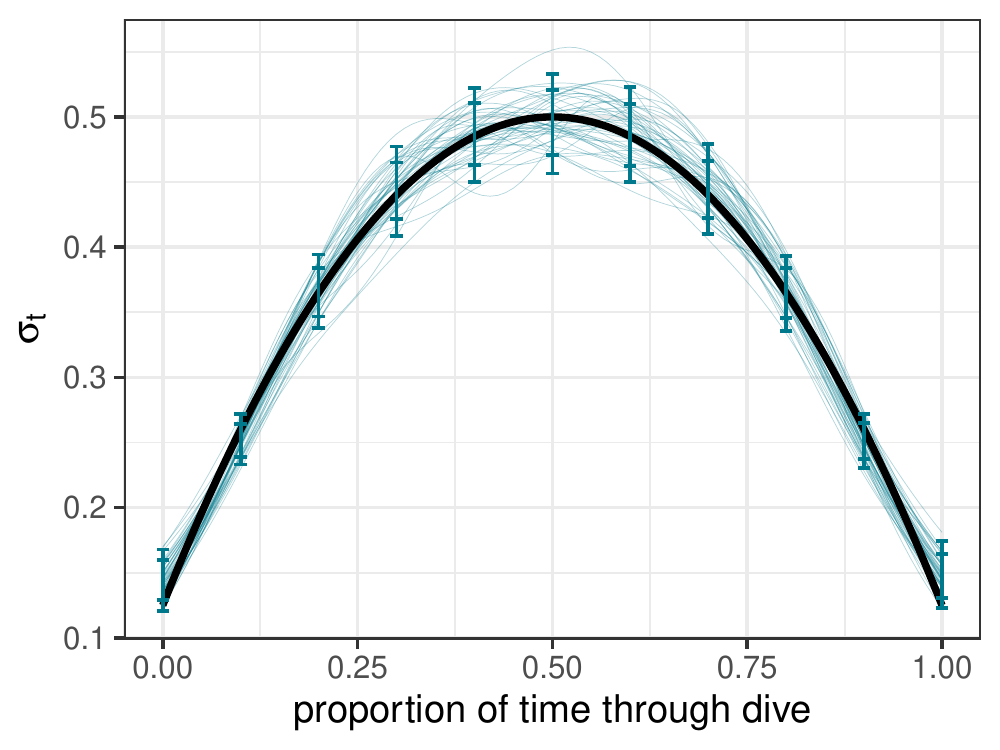}
  \includegraphics[width=0.49\textwidth]{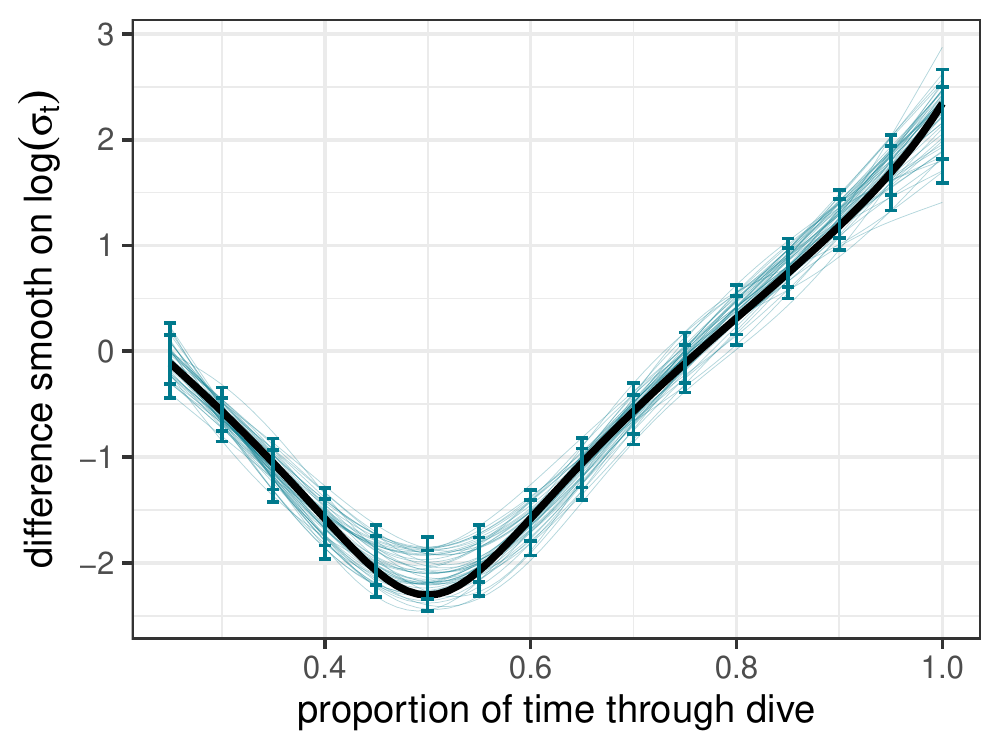}
  \caption{Estimated baseline model (left) and difference smooths (right) in simulation study. The thin blue lines are 50 randomly-selected estimated smooths (out of 2000), the vertical blue segments show the 2.5\%, 10\%, 90\%, and 97.5\% pointwise quantiles of the 2000 estimated smooths, and the thick black line is the true function used in the simulation.}
  \label{fig:sim_diffsmooth}
\end{figure}

For each simulated data set, we fitted a varying-coefficient Brownian motion with a difference smooth on the diffusion parameter, to capture the discrepancy between the dynamics of the process during baseline and response phases, similar to Equation \ref{eqn:par_bm2}. We repeated this procedure 2000 times, and the results are shown in Figure \ref{fig:sim_diffsmooth}. The results suggest that both the baseline model and the deviation from baseline (difference smooth) were well estimated. In particular, both the smoothness and the shape of the true functions used to simulate were captured well by the fitted splines. We also used these simulations to check the coverage of the simultaneous confidence intervals for the difference smooth. We found that the 95\% confidence band included the entire true function in 95.6\% of the simulation runs, indicating good coverage.

\section{Beaked whale case study}
\label{sec:whales}

For all analyses, we used the R package smoothSDE, available on Github at \url{https://github.com/TheoMichelot/smoothSDE} \citep{michelot2021}. The appendices include additional details about data and implementation.

\subsection{Horizontal avoidance}
\label{sec:horizontal}

We analysed the Argos trajectory described in Section \ref{sec:data_sat} using the model for horizontal avoidance described in Section \ref{sec:methods_horizontal}, embedded in a state-space formulation (Section \ref{sec:SSM}) to account for measurement error. The error ellipses from the satellite tag were used to create a covariance matrix for the animal's location at each time of observation, which was then used to account for measurement uncertainty in the likelihood ($\bm{H}_i$ in Equation \ref{eq:SSM_obs}). The data set was complemented with goniometer observations, which were more precise but only covered a short time period. The data were highly irregular, with intervals ranging from a few seconds to over a day, but the continuous-time approach could still be applied directly. 

\begin{figure}[htbp]
  \centering
  \includegraphics[width=\textwidth]{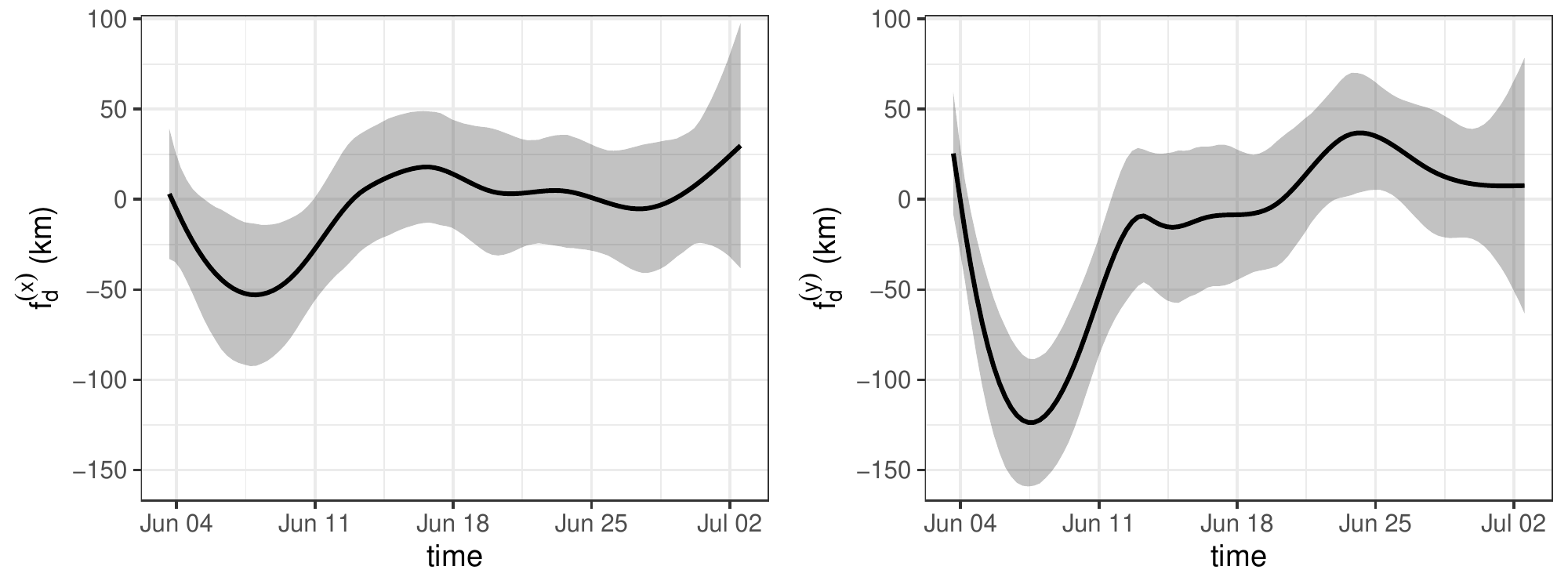}
  \caption{Estimated difference smooths for beaked whale centre of attraction, in the $x$ coordinate (left) and $y$ coordinate (right), as functions of time (after start of exposure). Grey areas are 95\% simultaneous confidence intervals. Deviations from zero suggest drift away from the whale's baseline centre of attraction (i.e., horizontal avoidance).}
  \label{fig:zc069_mu_est}
\end{figure}

Figure \ref{fig:zc069_mu_est} shows the time-varying centre of attraction parameters in each dimension (Easting and Northing), between the start of exposure and the end of the study. The negative drift in both dimensions suggests that the centre of attraction deviated towards the South-West, for about a week following the start of exposure in the afternoon of June 3rd (June 4th to June 11th). The maximum deviation in the first coordinate was estimated to be about 50km to the West, and the maximum deviation in the second coordinate was about 125km to the South. After June 12th, the whale seemed to revert to its baseline centre of attraction for the remainder of the study period.

This analytical method has provided evidence of unusual horizontal movement following exposure, and the consequences of such a large-scale movement (in both time and space) for the individual whale are not fully understood. These results should be interpreted in the context of all other available information (e.g., dive data, visual observations, biological knowledge, expert judgement) to inform a conclusion about whether the behaviour change was a response to the sonar exposure, the severity of the response, and the possible consequences (which is beyond the scope of this analysis). % For example, we may assume some energetic cost associated with moving such a distance, but it is unknown whether, and for how long, the animal stopped feeding. Even if it continued feeding, the animal may have moved to a suboptimal region for foraging.

\subsection{Unusual diving behaviour}
\label{sec:vertical}

We used the model for depth described in Section \ref{sec:methods_vertical} to analyse time series of depth collected from five beaked whales. The data set comprised 13 deep dives, and included two controlled exposure experiments. We downsampled depth to a 15-sec time resolution to reduce the computational cost of model fitting while retaining information about fine-scale behaviour.

\begin{figure}[htbp]
  \centering
  \includegraphics[width=\textwidth]{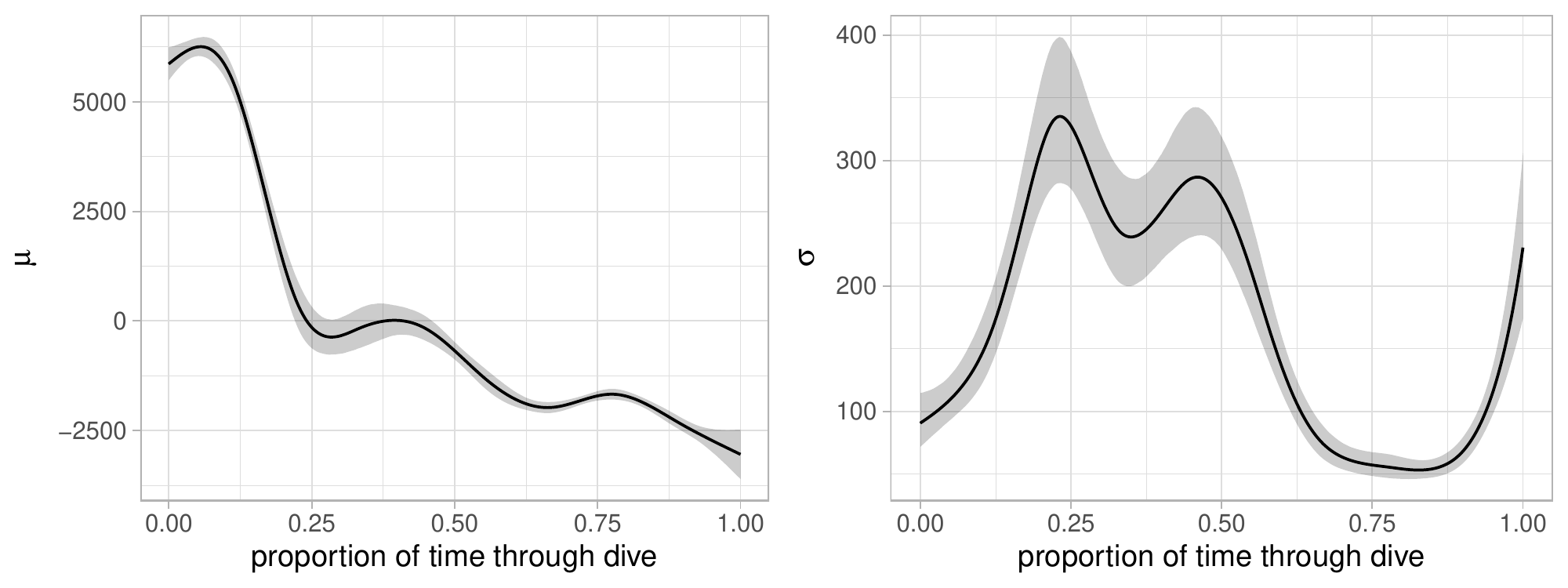}
  \caption{Estimated baseline parameters in the study of beaked whale diving behaviour: the drift $\mu$ is the expected change (left), and the diffusion $\sigma$ is the variability (right). The shaded areas show 95\% pointwise confidence bands.}
  \label{fig:dtag_baseline}
\end{figure}

The estimated parameters for the baseline model are presented in Figure \ref{fig:dtag_baseline} as functions of the proportion of time through a dive. The drift parameter, which measures the mean direction of change, was positive during the descent phase of the dive (because the depth increases), then close to zero during the bottom phase, and negative during the ascent phase (when the depth decreases and the animal returns to the surface). The diffusion parameter was highest during the bottom phase of the dive, suggesting high variability due to active foraging behaviour.

We used posterior predictive checks to evaluate whether the chosen Brownian motion was an appropriate model of baseline diving behaviour. We applied the procedure described in Section \ref{sec:check} to compare the true data to simulations from the model. We simulated 1000 baseline dives, and compared them to the observed baseline dives based on the following metrics:
\begin{itemize}
\item proportion of time ascending, measured by proportion of time steps where $D_{i+1} > D_i + 10$ (i.e., depth increases by more than 10 metres over a 15-sec interval);
\item proportion of time descending, i.e., proportion of time steps where $D_{i+1} < D_i - 10$;
\item maximum depth;
\item proportion of time spent deeper than 500m;
\item proportion of time spent deeper than 1000m;
\item persistence in vertical direction of movement, i.e., proportion of consecutive pairs of time steps where direction of movement remains the same (either ascending or descending).
\end{itemize}

\begin{figure}[htbp]
  \centering
  \includegraphics[width=0.9\textwidth]{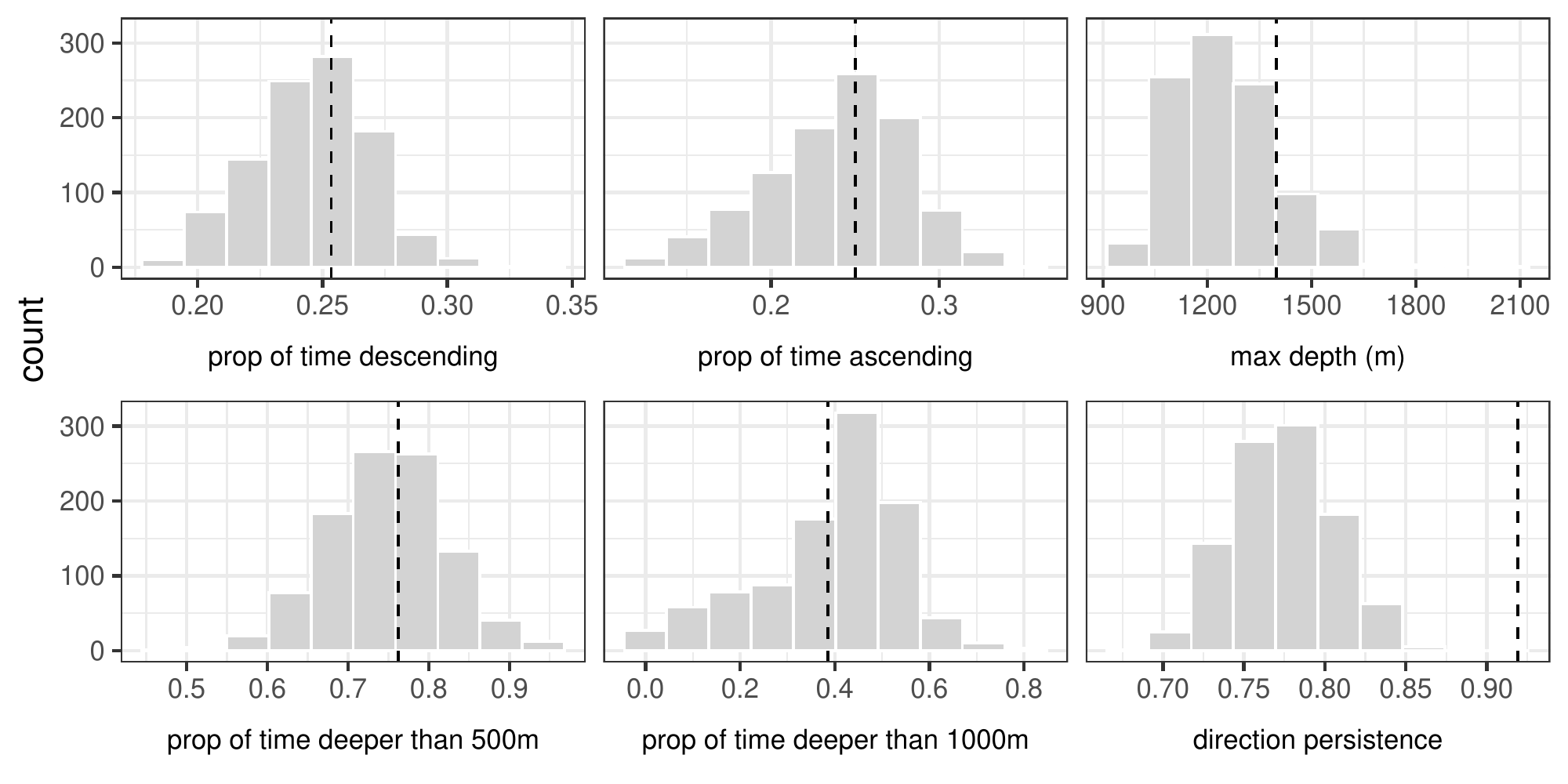}
  \caption{Goodness-of-fit for baseline model of beaked whale diving behaviour. Each plot contrasts the observed mean of a metric for baseline dives (vertical dotted line), and a histogram of values obtained from 1000 simulated dives. An observed value in the tails of the distribution suggests lack of fit. The metrics are described in the text.}
  \label{fig:dtag_gof}
\end{figure}

These metrics were chosen to assess how well the fitted model captured the shape of baseline dives. The results are shown in Figure \ref{fig:dtag_gof}. For the first five metrics, the mean observed value lay well within the distribution of simulation values, suggesting that features related to the overall shape of dives were well captured by the model. The observed data, however, displayed stronger directional persistence than the simulated dives. This illustrates the limited ability of Brownian motion to capture persistent movement, as there is no built-in mechanism to create directional autocorrelation. It is worth noting that the directional persistence of the simulated dives was between 0.68 and 0.86, which is much higher than the value of 0.5 expected under a simple random walk with no time-varying parameters. This is because some correlation in direction is induced by the model used for the drift parameter of the process.

\begin{figure}[htbp]
  \centering
  \includegraphics[width=0.49\textwidth]{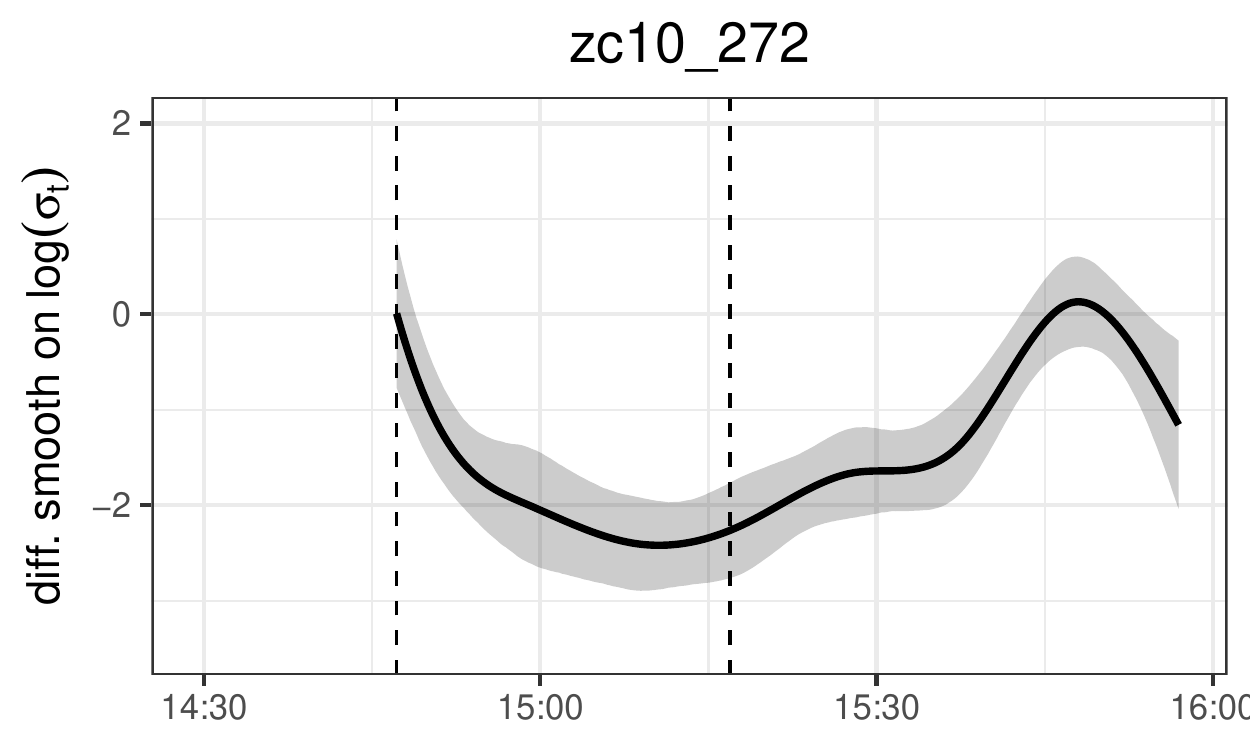}
  \includegraphics[width=0.49\textwidth]{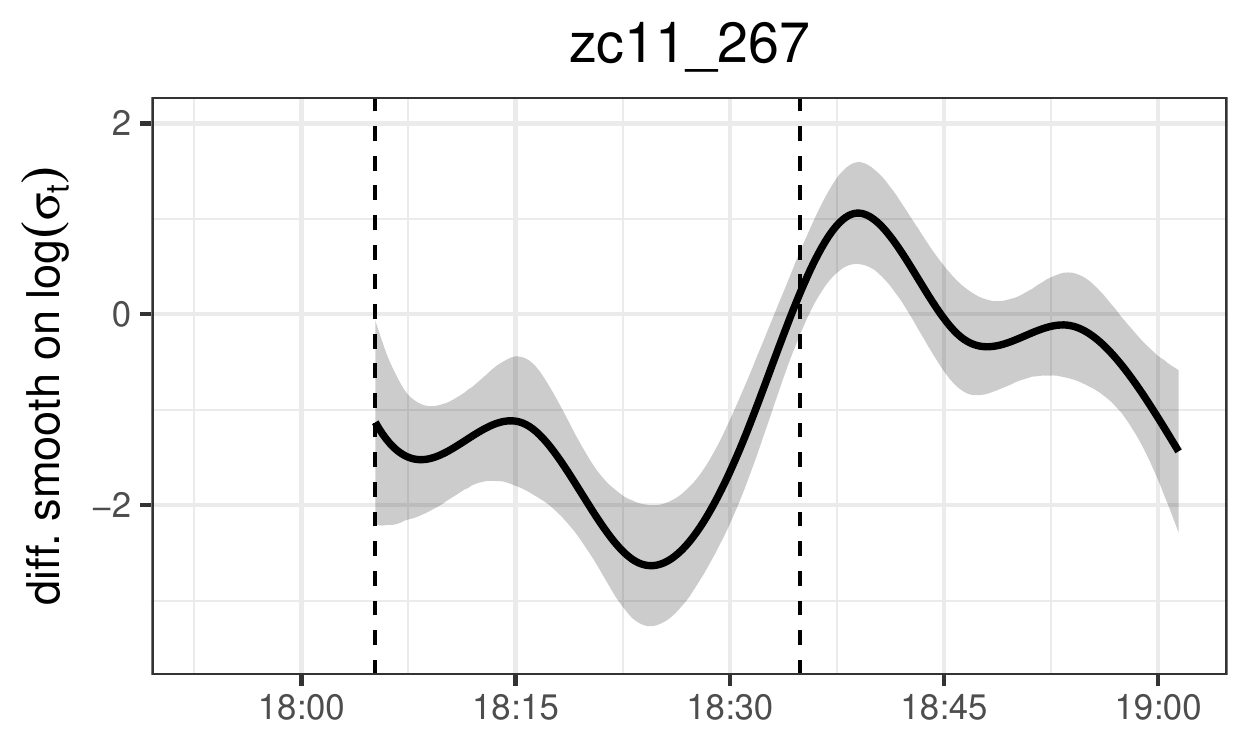}\\
  \includegraphics[width=0.49\textwidth]{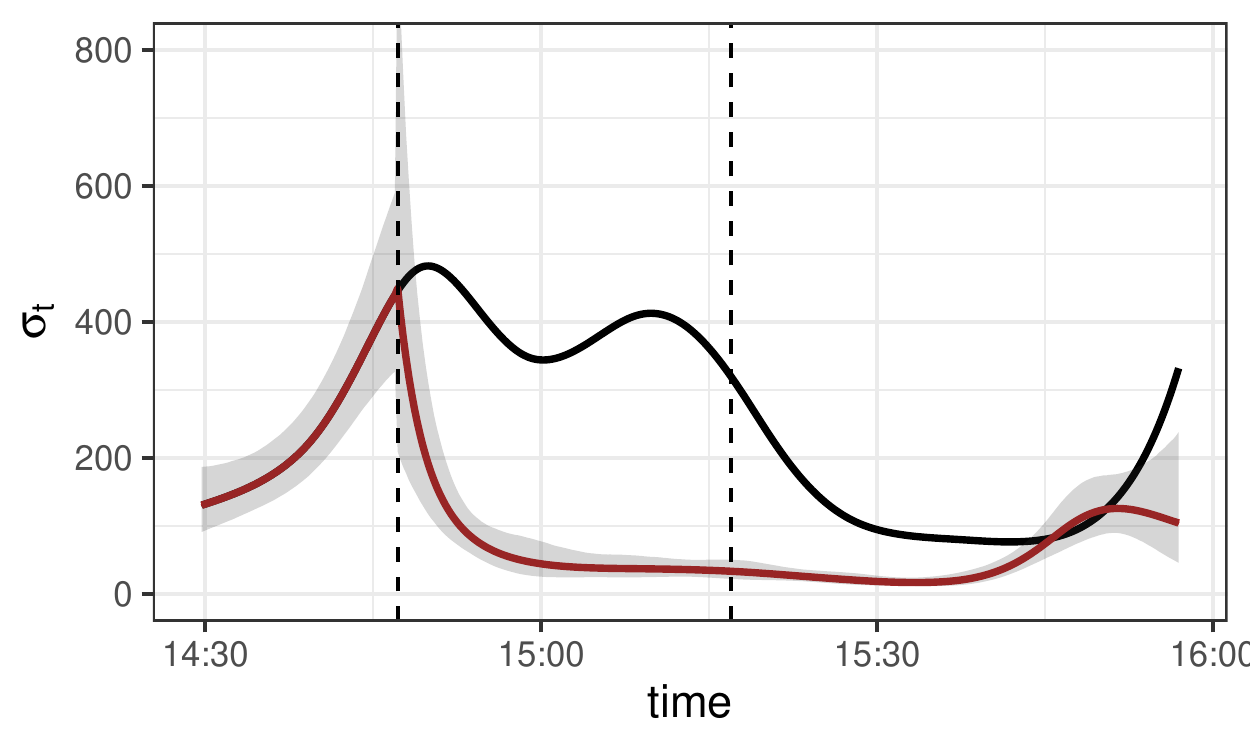}
  \includegraphics[width=0.49\textwidth]{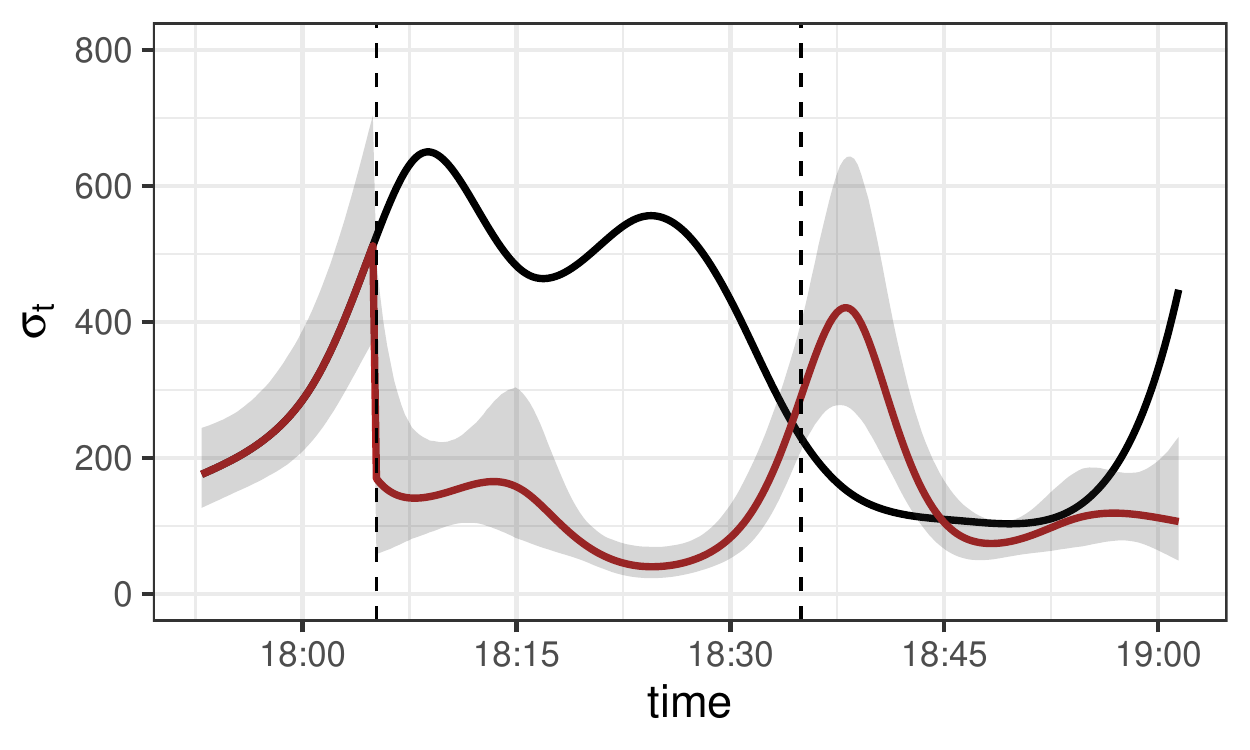}
  \caption{Results of analysis of beaked whale diving behaviour. Top row: difference smooths for the diffusion parameter $\sigma_t$ (on log scale) during the two exposed dives, with 95\% simultaneous confidence bands. Deviation from zero suggests behaviour inconsistent with baseline model. Bottom row: black lines show the baseline estimate for $\sigma_t$, and the red lines show the response model (including difference smooths), with 95\% simultaneous confidence bands. The baseline model is not identical for the two dives (and also different from Figure \ref{fig:dtag_baseline}) due to the dive-specific random intercept. In all plots, the vertical dashed lines mark the start and end of sound exposure.}
  \label{fig:dtag_diffsmooth}
\end{figure}

In this model, behavioural responses were modelled using difference smooths, i.e., functions capturing deviations from the baseline model during exposed dives. One difference smooth was estimated for the diffusion parameter $\sigma$ for each of the two exposed dives, to estimate the difference in the depth variability compared to a typical baseline dive. Figure \ref{fig:dtag_diffsmooth} shows the estimated smooths with confidence bands. Both curves display large departures from zero after the start of exposure, indicating deviations from baseline behaviour. In the first exposed dive (``zc10\_272''), the most noticeable pattern is that the diffusion parameter was much lower than normal during the bottom phase. This decreased variability in depth reflects a reduced level of activity, which suggests that this animal was not searching or chasing prey, i.e., this was not a foraging dive. In the second exposed dive (``zc11\_267''), $\sigma_t$ was also low during the middle part of the dive, which was followed by a period of unusually high vertical activity during the third quarter of the dive. This period coincides with the bottom phase of this dive, which was delayed because the animal appeared to be engaged in a bout of shallow diving, but conducted a deep dive after the start of exposure (see Figure S2 in Appendix A).

% Similarly to horizontal avoidance, these behavioural changes are likely to have energetic and physiological consequences for the individual. Deep dives are generally associated with foraging, and so conducting a deep dive without feeding presumably incurs a cost to the animal with no associated benefit. In addition, switching from a planned shallow dive to a deep dive following exposure may push the animal to its physiological limits as shallow dives are important for recovery from deep dives.

\section{Discussion}
\label{sec:discussion}

Varying-coefficient SDEs are versatile models to study the effects of covariates on the dynamics of temporal processes. Here, we have demonstrated their utility to estimate behavioural changes in cetaceans, including changes in the pattern of diving behaviour, and horizontal avoidance. This method stands in contrast with previous statistical approaches applied in this context, where data have often been summarised at the time scale of dives \citep[e.g., dive duration, maximum depth;][]{deruiter2013, deruiter2017}. Our analysis of DTag data at a fine time scale is one of the first attempts to describe detailed dynamics of within-dive behaviour, and contrast them between baseline and exposed dives. Conclusions about whether an individual exhibited a behavioural response to a sonar exposure event often requires multiple lines of evidence to be examined and expert biological knowledge applied to interpret the results from models. In particular, it is experts that are required to evaluate responses in terms of their severity and possible effect on the vital rates of individuals \citep{southall2008, miller2012}. We have shown that varying-coefficient SDEs can provide such lines of evidence for different types of behaviour (horizontal movement, diving) and data recorded at different resolutions from different types of telemetry devices. Critically, our method provides fine-scale detail about the nature of the response, in the context of relevant covariates, as well as the duration of the response. This is valuable additional information beyond the identification of a behavioural change point. Some other analytical methods, such as Mahalanobis distance, might not be able to identify behavioural responses that are within the repertoire of baseline behaviour, even if their occurrence is unusual or unexpected given what the animal was doing at the point of exposure.

SDEs have recently been proposed to model group movement of animals \citep{niu2016, milner2021}. That approach could be extended to allow for time-varying dynamics and, using the methods presented in this paper, it could be used to estimate behavioural responses for multiple individual animals. This model would not be limited to individual-specific responses, but could also capture changes in the interactions between individuals (e.g., group breaking off after disturbance). 

The examples that we presented illustrate a general framework to analyse behavioural responses from telemetry data. The first step is to specify a model for baseline periods, typically in terms of spatial or temporal covariates of interest (e.g., time of day, habitat variable). Then, additional terms can be added in the model for the SDE parameters, to capture deviations from baseline during exposure phases. Difference smooths are powerful for this purpose, as they explicitly model the difference in a smooth relationship between levels of a categorical variable (which could for example represent pre- and post-disturbance). We showcased how difference smooths can be interpreted in terms of behavioural response, for two different applications. In particular, simultaneous confidence bands are useful to compare the deviation from baseline to the zero function. Although we checked the coverage of these confidence intervals in simulations, we note that this method might lead to a large rate of false positives in real data applications if the model assumptions are violated (e.g., if the baseline model does not adequately capture heterogeneity in baseline data). For this reason, the shape and amplitude of the difference smooth should be inspected as part of the interpretation, rather than merely whether it clearly differs from zero. A wide range of varying-coefficient SDEs can be implemented using the smoothSDE R package, and we anticipate that these methods will be a key tool to investigate the potential impact of disturbance, such as sonar, on individuals and populations.

\subsection*{Acknowledgements}

\textit{\small We are very grateful to Rob Schick, Will Cioffi, Alan Gelfand, Josh Hewitt, Stacy DeRuiter, and Brandon Southall for discussions about the data and models. TM, RG, CH, and LT were funded by the US office of Naval Research, Grant N000141812807. The data from four of the five DTags were collected as part of the SOCAL-BRS project, primarily funded by the US Navy’s Chief of Naval Operations Environmental Readiness Division and subsequently by the US Navy's Living Marine Resources Program. Additional support for environmental sampling and logistics was also provided by the Office of Naval Research, Marine Mammal Program. All research activities for that study were authorized and conducted under US National Marine Fisheries Service permit 14534; Channel Islands National Marine Sanctuary permit 2010-004; US Department of Defense Bureau of Medicine and Surgery authorization; a federal consistency determination by the California Coastal Commission; and numerous institutional animal care and use committee authorizations. The data from the satellite tag and one of the DTags were collected as part of the Atlantic BRS project under National Marine Fisheries Service scientific research permit numbers 17086 and 20605 to Robin W.\ Baird. The tagging protocol was approved by the Institutional Animal Care and Use Committee at Cascadia Research Collective. This work was supported by the US Fleet Forces Command through the Naval Facilities Engineering Command Atlantic under Contract No.\ N62470-15-D-8006, Task Order 50, Issued to HDR, Inc. We thank all members of the field teams involved in both the SOCAL and Atlantic BRS projects.}

\bibliographystyle{apalike}
\bibliography{refs.bib}

\begin{thebibliography}{}

\bibitem[Anderson-Sprecher and Ledolter, 1991]{anderson1991}
Anderson-Sprecher, R. and Ledolter, J. (1991).
\newblock State-space analysis of wildlife telemetry data.
\newblock {\em Journal of the American Statistical Association},
  86(415):596--602.

\bibitem[Andrews et~al., 2008]{andrews2008}
Andrews, R.~D., Pitman, R.~L., and Ballance, L.~T. (2008).
\newblock Satellite tracking reveals distinct movement patterns for {Type} {B}
  and {Type} {C} killer whales in the southern {Ross} {Sea}, {Antarctica}.
\newblock {\em Polar Biology}, 31(12):1461--1468.

\bibitem[Cioffi et~al., 2022]{cioffi2022}
Cioffi, W.~R., Quick, N.~J., Swaim, Z.~T., Foley, H.~J., Waples, D.~M.,
  Webster, D.~L., Baird, R.~W., Southall, B.~L., Nowacek, D.~P., and Read,
  A.~J. (2022).
\newblock Trade-offs in telemetry tag programming for deep-diving cetaceans:
  data longevity, resolution, and continuity.
\newblock {\em bioRxiv}.

\bibitem[DeRuiter et~al., 2017]{deruiter2017}
DeRuiter, S.~L., Langrock, R., Skirbutas, T., Goldbogen, J.~A., Calambokidis,
  J., Friedlaender, A.~S., and Southall, B.~L. (2017).
\newblock A multivariate mixed hidden {Markov} model for blue whale behaviour
  and responses to sound exposure.
\newblock {\em The Annals of Applied Statistics}, 11(1):362--392.

\bibitem[DeRuiter et~al., 2013]{deruiter2013}
DeRuiter, S.~L., Southall, B.~L., Calambokidis, J., Zimmer, W.~M., Sadykova,
  D., Falcone, E.~A., Friedlaender, A.~S., Joseph, J.~E., Moretti, D., Schorr,
  G.~S., et~al. (2013).
\newblock First direct measurements of behavioural responses by {Cuvier's}
  beaked whales to mid-frequency active sonar.
\newblock {\em Biology letters}, 9(4):20130223.

\bibitem[Dunn and Gipson, 1977]{dunn1977}
Dunn, J.~E. and Gipson, P.~S. (1977).
\newblock Analysis of radio telemetry data in studies of home range.
\newblock {\em Biometrics}, pages 85--101.

\bibitem[Durbin and Koopman, 2012]{durbin2012}
Durbin, J. and Koopman, S.~J. (2012).
\newblock {\em Time series analysis by state space methods}.
\newblock Oxford university press.

\bibitem[Foley et~al., 2021]{foley2021}
Foley, H.~J., Pacifici, K., Baird, R.~W., Webster, D.~L., Swaim, Z.~T., and
  Read, A.~J. (2021).
\newblock Residency and movement patterns of {Cuvier}’s beaked whales
  {Ziphius} cavirostris off {Cape} {Hatteras}, {North} {Carolina}, {USA}.
\newblock {\em Marine Ecology Progress Series}, 660:203--216.

\bibitem[Johnson et~al., 2008]{johnson2008}
Johnson, D.~S., London, J.~M., Lea, M.-A., and Durban, J.~W. (2008).
\newblock Continuous-time correlated random walk model for animal telemetry
  data.
\newblock {\em Ecology}, 89(5):1208--1215.

\bibitem[Johnson and Tyack, 2003]{johnson2003}
Johnson, M.~P. and Tyack, P.~L. (2003).
\newblock A digital acoustic recording tag for measuring the response of wild
  marine mammals to sound.
\newblock {\em IEEE journal of oceanic engineering}, 28(1):3--12.

\bibitem[Jones-Todd et~al., 2022]{jones2022}
Jones-Todd, C.~M., Pirotta, E., Durban, J.~W., Claridge, D.~E., Baird, R.~W.,
  Falcone, E.~A., Schorr, G.~S., Watwood, S., and Thomas, L. (2022).
\newblock Discrete-space continuous-time models of marine mammal exposure to
  {Navy} sonar.
\newblock {\em Ecological Applications}, 32(1):e02475.

\bibitem[Jonsen et~al., 2003]{jonsen2003}
Jonsen, I.~D., Myers, R.~A., and Mills~Flemming, J. (2003).
\newblock Meta-analysis of animal movement using state-space models.
\newblock {\em Ecology}, 84(11):3055--3063.

\bibitem[Kristensen et~al., 2016]{kristensen2015}
Kristensen, K., Nielsen, A., Berg, C.~W., Skaug, H., and Bell, B.~M. (2016).
\newblock {TMB}: Automatic differentiation and {L}aplace approximation.
\newblock {\em Journal of Statistical Software}, 70(5):1--21.

\bibitem[Marra and Wood, 2012]{marra2012}
Marra, G. and Wood, S.~N. (2012).
\newblock Coverage properties of confidence intervals for generalized additive
  model components.
\newblock {\em Scandinavian Journal of Statistics}, 39(1):53--74.

\bibitem[McClintock et~al., 2015]{mcclintock2015}
McClintock, B.~T., London, J.~M., Cameron, M.~F., and Boveng, P.~L. (2015).
\newblock Modelling animal movement using the {Argos} satellite telemetry
  location error ellipse.
\newblock {\em Methods in Ecology and Evolution}, 6(3):266--277.

\bibitem[Michelot et~al., 2021]{michelot2021}
Michelot, T., Glennie, R., Harris, C., and Thomas, L. (2021).
\newblock Varying-coefficient stochastic differential equations with
  applications in ecology.
\newblock {\em Journal of Agricultural, Biological and Environmental
  Statistics}, 26:446--463.

\bibitem[Michelot et~al., 2019]{michelot2019langevin}
Michelot, T., Gloaguen, P., Blackwell, P.~G., and {\'E}tienne, M.-P. (2019).
\newblock The {Langevin} diffusion as a continuous-time model of animal
  movement and habitat selection.
\newblock {\em Methods in Ecology and Evolution}, 10(11):1894--1907.

\bibitem[Miller, 2019]{miller2019}
Miller, D.~L. (2019).
\newblock Bayesian views of generalized additive modelling.
\newblock {\em arXiv preprint arXiv:1902.01330}.

\bibitem[Miller et~al., 2012]{miller2012}
Miller, P.~J., Kvadsheim, P.~H., Lam, F.-P.~A., Wensveen, P.~J., Antunes, R.,
  Alves, A.~C., Visser, F., Kleivane, L., Tyack, P.~L., and Sivle, L.~D.
  (2012).
\newblock The severity of behavioral changes observed during experimental
  exposures of killer ({Orcinus} orca), long-finned pilot ({Globicephala}
  melas), and sperm ({Physeter} macrocephalus) whales to naval sonar.
\newblock {\em Aquatic Mammals}, 38(4):362.

\bibitem[Milner et~al., 2021]{milner2021}
Milner, J.~E., Blackwell, P.~G., and Niu, M. (2021).
\newblock Modelling and inference for the movement of interacting animals.
\newblock {\em Methods in Ecology and Evolution}, 12(1):54--69.

\bibitem[Niu et~al., 2016]{niu2016}
Niu, M., Blackwell, P.~G., and Skarin, A. (2016).
\newblock Modeling interdependent animal movement in continuous time.
\newblock {\em Biometrics}, 72(2):315--324.

\bibitem[Pozdnyakov et~al., 2014]{pozdnyakov2014}
Pozdnyakov, V., Meyer, T., Wang, Y.-B., and Yan, J. (2014).
\newblock On modeling animal movements using {Brownian} motion with measurement
  error.
\newblock {\em Ecology}, 95(2):247--253.

\bibitem[Rigby and Stasinopoulos, 2005]{rigby2005}
Rigby, R.~A. and Stasinopoulos, D.~M. (2005).
\newblock Generalized additive models for location, scale and shape.
\newblock {\em Journal of the Royal Statistical Society: Series C (Applied
  Statistics)}, 54(3):507--554.

\bibitem[Ruppert et~al., 2003]{ruppert2003}
Ruppert, D., Wand, M.~P., and Carroll, R.~J. (2003).
\newblock {\em Semiparametric regression}.
\newblock Cambridge university press.

\bibitem[Shearer et~al., 2019]{shearer2019}
Shearer, J.~M., Quick, N.~J., Cioffi, W.~R., Baird, R.~W., Webster, D.~L.,
  Foley, H.~J., Swaim, Z.~T., Waples, D.~M., Bell, J.~T., and Read, A.~J.
  (2019).
\newblock Diving behaviour of {Cuvier}'s beaked whales ({Ziphius} cavirostris)
  off {Cape} {Hatteras}, {North} {Carolina}.
\newblock {\em Royal Society Open Science}, 6(2):181728.

\bibitem[Southall et~al., 2020]{southall2020}
Southall, B.~L., Bowers, M., Cioffi, W., Foley, H., Harris, C., Joseph, J.,
  Quick, N., Margolina, T., Nowacek, D., Read, A.~J., Schick, R., Swaim, Z.~T.,
  Waples, D., and Webster, D.~L. (2020).
\newblock Atlantic behavioral response study ({BRS}): 2019 annual progress
  report.
\newblock Project report. Prepared for U.S. Fleet Forces Command. Submitted to
  Naval Facilities Engineering Command Atlantic, Norfolk, Virginia, under
  Contract No. N62470-15-D-8006, Task Order 19F4029, issued to HDR Inc.,
  Virginia Beach, Virginia. May 2020.

\bibitem[Southall et~al., 2008]{southall2008}
Southall, B.~L., Bowles, A.~E., Ellison, W.~T., Finneran, J.~J., Gentry, R.~L.,
  Jr., C. R.~G., Kastak, D., Ketten, D.~R., Miller, J.~H., Nachtigall, P.~E.,
  Richardson, W.~J., Thomas, J.~A., and Tyack, P.~L. (2008).
\newblock Marine mammal noise-exposure criteria: initial scientific
  recommendations.
\newblock {\em Bioacoustics}, 17(1-3):273--275.

\bibitem[Southall et~al., 2019]{southall2019}
Southall, B.~L., Finneran, J.~J., Reichmuth, C., Nachtigall, P.~E., Ketten,
  D.~R., Bowles, A.~E., Ellison, W.~T., Nowacek, D.~P., and Tyack, P.~L.
  (2019).
\newblock Marine mammal noise exposure criteria: updated scientific
  recommendations for residual hearing effects.
\newblock {\em Aquatic Mammals}, 45(2).

\bibitem[Southall et~al., 2016]{southall2016experimental}
Southall, B.~L., Nowacek, D.~P., Miller, P.~J., and Tyack, P.~L. (2016).
\newblock Experimental field studies to measure behavioral responses of
  cetaceans to sonar.
\newblock {\em Endangered Species Research}, 31:293--315.

\bibitem[Stasinopoulos and Rigby, 2008]{stasinopoulos2008}
Stasinopoulos, D.~M. and Rigby, R.~A. (2008).
\newblock Generalized additive models for location scale and shape ({GAMLSS})
  in {R}.
\newblock {\em Journal of Statistical Software}, 23:1--46.

\bibitem[Stimpert et~al., 2014]{stimpert2014}
Stimpert, A., DeRuiter, S.~L., Southall, B., Moretti, D., Falcone, E.,
  Goldbogen, J., Friedlaender, A., Schorr, G., and Calambokidis, J. (2014).
\newblock Acoustic and foraging behavior of a {Baird}'s beaked whale,
  {Berardius} bairdii, exposed to simulated sonar.
\newblock {\em Scientific Reports}, 4(1):1--8.

\bibitem[Tyack et~al., 2011]{tyack2011}
Tyack, P.~L., Zimmer, W.~M., Moretti, D., Southall, B.~L., Claridge, D.~E.,
  Durban, J.~W., Clark, C.~W., D'Amico, A., DiMarzio, N., Jarvis, S., et~al.
  (2011).
\newblock Beaked whales respond to simulated and actual navy sonar.
\newblock {\em PloS one}, 6(3):e17009.

\end{thebibliography}

\newpage
\begin{appendix}
\renewcommand\thefigure{S\arabic{figure}}
\renewcommand\thetable{S\arabic{table}}
\setcounter{figure}{0}

\thispagestyle{empty}
\begin{center}
  \LARGE\bf
  Appendices for ``Continuous-time modelling of behavioural responses in animal movement''
\end{center}

\section{Data overview}

\begin{figure}[htbp]
  \centering
  \includegraphics[width=\textwidth]{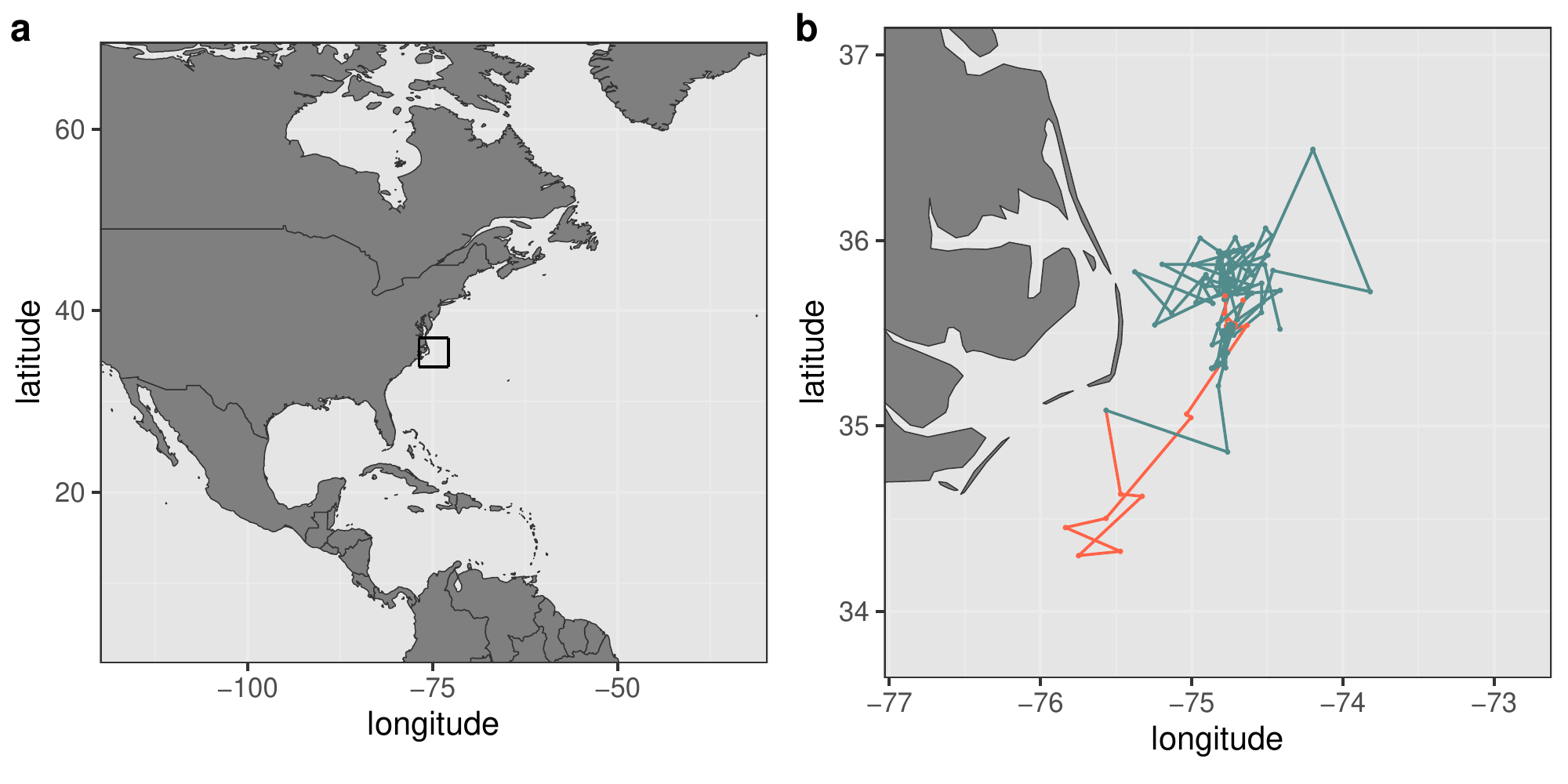}
  \caption{(a) The study region (enlarged in (b)) is shown as a rectangle on the East coast of the USA. (b) Beaked whale tracking data, obtained from a satellite tag and goniometer. Data collected during the week following exposure are shown in red.}
  \label{fig:data_sat}
\end{figure}

\paragraph{Satellite data} The satellite data came from the individual ``zc069'', and it included 77 Argos locations and 294 goniometer locations after preprocessing (i.e., 371 locations in total). The track covered a period of about 38 days between May 24th 2018 and July 2nd 2018, and sound exposure occured on June 3rd 2018. Figure \ref{fig:data_sat} shows a plot of the movement track.

\begin{figure}[htbp]
  \centering
  \includegraphics[width=\textwidth]{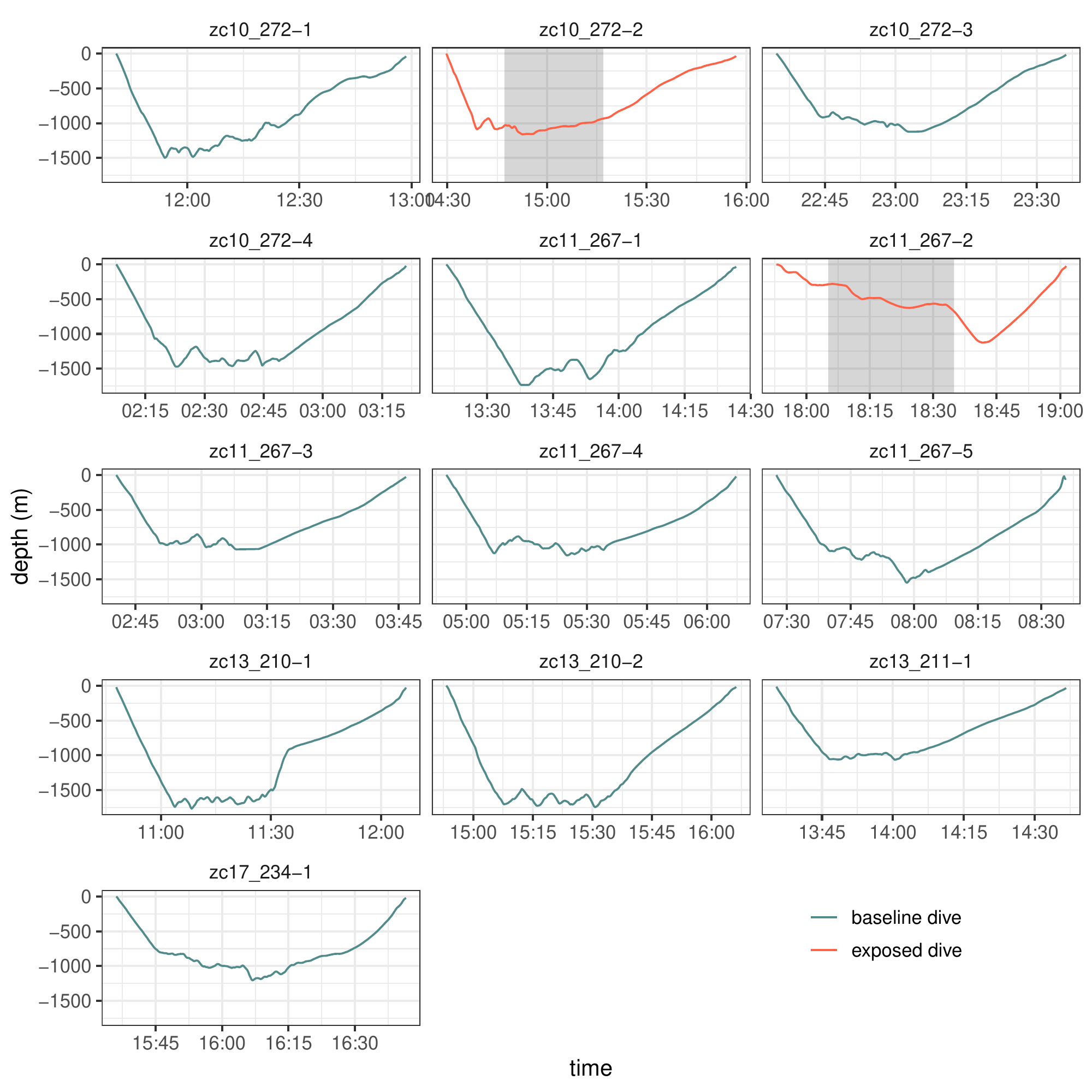}
  \caption{Depth of beaked whales over 13 deep dives, obtained from DTags. The blue lines are baseline dives, and the red lines are dives during which a sound exposure occured. For the two exposed dives, the period of exposure is shown as a shaded area.}
  \label{fig:data_dtag}
\end{figure}

\paragraph{DTag data} We summarise information about the DTag data in Table \ref{tab:data}, and times series plots of depth for all deep dives are shown in Figure \ref{fig:data_dtag}.

\begin{table}[htbp]
  \centering
  \caption{DTag data summary. For each whale, this gives the number of observations (at the 15-sec resolution), the number of deep dives used in the analysis, and whether the individual was exposed to sonar during the study.} 
  \begin{tabular}{cccc}
    \toprule
    Animal ID & Number of observations & Number of deep dives & Exposed? \\
    \midrule
    zc10\_272 & 1203 & 4 & yes \\
    zc11\_267 & 1369 & 5 & yes \\
    zc13\_210 & 610 & 2 & no \\
    zc13\_211 & 246 & 1 & no \\
    zc17\_234 & 262 & 1 & no \\
    \bottomrule
  \end{tabular}
  \label{tab:data}
\end{table}

\section{Implementation details}

We used the R package smoothSDE to implement all analyses (\url{https://github.com/TheoMichelot/smoothSDE}). Here, we summarise the model formulations used for all varying-coefficient SDEs fitted in the paper, and the initial parameter values used for the likelihood optimisation.

\subsection{Simulation study}

The data simulation procedure is described in the main text. For the estimation, we fitted a Brownian motion model with constant drift, and with the following formula on the diffusion parameter $\sigma_t$ (or, more precisely, on $\log(\sigma_t)$),

\hspace{0.1\linewidth}
\begin{minipage}{0.5\linewidth}
\begin{verbatim}
~ expo + 
  s(diveprop, k = 10, bs = "ts") +
  s(diveprop, by = expo, k = 10, bs = "ts")
\end{verbatim}
\end{minipage}

This uses the syntax from the R package mgcv, and we refer to its documentation for details. In brief, the first term includes a separate intercept for exposed and non-exposed dives, the second term is a thin-plate regression spline (with basis dimension = 10) for baseline, and the third term is a difference smooth for exposed dives, also modelled with a thin-plate regression spline (with basis dimension = 10).

For model fitting, we needed to choose initial values for all model parameters. For the drift parameter, we selected an initial value of zero. For the diffusion parameter, we initialised the spline basis coefficients to zero (i.e., no covariate effects), and we initialised the intercept to $\log(0.3)$.

\subsection{Horizontal avoidance analysis}

We fitted an isotropic Ornstein-Uhlenbeck process to the bivariate location $\bm{Z}_t = (Z_t^x, Z_t^y)$ of the animal; i.e., it was defined by the SDEs
\begin{equation*}
  \begin{cases}
    dZ_t^x = b(a_t^x - Z_t) dt + \sigma dW_t\\
    dZ_t^y = b(a_t^y - Z_t) dt + \sigma dW_t
  \end{cases}
\end{equation*}
with constant parameters $b$ and $\sigma$. Each coordinate of the time-varying centre of attraction $\bm{a}_t = (a_t^x, a_t^y)$ was estimated separately with the formula

\hspace{0.1\linewidth}
\begin{minipage}{0.5\linewidth}
\begin{verbatim}
~ expo +
  s(t_expo, by = expo, k = 20, bs = "cs")
\end{verbatim}
\end{minipage}

This indicates that the centre of attraction was assumed to be constant during baseline behaviour (and modelled by the implicit intercept in the formula), and we used a cubic spline (basis dimension = 20) for the difference smooth after start of exposure.

In smoothSDE, the Ornstein-Uhlenbeck process is specified in terms of the parameters $\tau = 1/b$ and $\kappa = \sigma^2/(2b)$, rather than $b$ and $\sigma$ directly, because the former often have a more natural interpretation. Indeed, $\tau$ is a measure of the time scale of autocorrelation of the process, and $\kappa$ is the variance of the stationary distribution of the process.

For starting values, we initialised all basis coefficients to zero (i.e., no effect of exposure), and we initialised the baseline parameters based on visual inspection of the data. Specifically, the baseline centre of attraction $\bm{a}$ was roughly chosen as the mean of the observed locations during the pre-exposure phase of the study. Then, $\tau$ was initialised to 10 hours, and $\kappa$ to 1000 (i.e., the standard deviation of the stationary distribution of the process was initialised to $\sqrt{\kappa} = \sqrt{1000} \approx 32$ km).

\subsection{Diving behaviour analysis}

We fitted a varying-coefficient Brownian motion, with the following formula for the drift parameter:

\hspace{0.1\linewidth}
\begin{minipage}{0.5\linewidth}
\begin{verbatim}
~ s(diveprop, k = 10, bs = "cs")
\end{verbatim}
\end{minipage}

That is, the drift was modelled as a smooth function of proportion of time through dive, using a cubic spline with basis dimension 10. The diffusion parameter was modelled with the formula

\hspace{0.1\linewidth}
\begin{minipage}{0.5\linewidth}
\begin{verbatim}
~ s(ID, bs = "re") +
  s(diveprop, k = 10, bs = "cs") +
  s(diveprop, by = exposed_dive, k = 10, bs = "cs")
\end{verbatim}
\end{minipage}

The first term is an iid normal random intercept for the dive, the second term is the baseline model (cubic spline with basis dimension 10), and the third term is the difference smooth to capture deviation from baseline in exposed dives (cubic spline with basis dimension 10).

We initialised all basis coefficients and random effects to zero (i.e., no covariate effects), and initialised the intercept parameters so that the starting value for the drift was zero, and the starting value for the diffusion parameter was 100.

\end{appendix}

\end{document}